\documentclass[twocolumn,superscriptaddress,pra]{revtex4}

\usepackage{graphicx}
\usepackage{delarray}
\usepackage{amsmath}
\usepackage{bm}

\newcommand{\ket}[1]{| { #1} \rangle}

\begin{document}
\title{All photonic quantum repeaters}

\author{Koji Azuma}
\email{azuma.koji@lab.ntt.co.jp}
\affiliation{NTT Basic Research Laboratories, NTT Corporation, 3-1 Morinosato Wakamiya, Atsugi, Kanagawa 243-0198, Japan}

\author{Kiyoshi Tamaki}
\affiliation{NTT Basic Research Laboratories, NTT Corporation, 3-1 Morinosato Wakamiya, Atsugi, Kanagawa 243-0198, Japan}

\author{Hoi-Kwong Lo}
\affiliation{Center for Quantum Information and Quantum Control (CQIQC),
Department of Physics and Department of Electrical \& Computer Engineering,
University of Toronto, Toronto, Ontario, M5S 3G4, Canada}

\date{\today}

%%%%%%%%%%%%%%%%%%%%%%%%%%%% abstract %%%%%%%%%%%%%%%%%%%%%%%%%%%%%

\begin{abstract}
Quantum communication holds promise for unconditionally secure transmission of secret messages and
faithful transfer of unknown quantum states. Photons appear to be the medium of choice for quantum
communication. Owing to photon losses, robust quantum communication over long lossy channels
requires quantum repeaters. It is widely believed that a necessary and highly demanding requirement
for quantum repeaters is the existence of matter quantum memories
at the repeater nodes. Here we show that such a requirement is, in fact, unnecessary by introducing the
concept of all photonic quantum repeaters based on flying qubits. As an example of the realization of this
concept, we present a protocol based on photonic cluster state machine guns and a loss-tolerant
measurement equipped with local high-speed active feedforwards. We show that, with such an all photonic
quantum repeater, the communication efficiency still scales polynomially with the channel distance.
Our result paves a new route toward quantum repeaters with efficient single-photon sources rather than matter quantum memories.
%\pacs{03.67.Hk, 03.67.Bg, 03.65.Ud, 03.67.Mn}
\end{abstract}
\maketitle

%%%%%%%%%%%%%%%%%%%%%%%%%%%%%%% intro %%%%%%%%%%%%%%%%%%%%%%%%%%%%%

\section{Introduction}
Quantum communication not only opens up opportunities for secure communications \cite{BB84,E91} and the
teleportation of quantum states \cite{B93}, but also is an important ingredient of the quantum internet \cite{K08} which
enables the distribution of entanglement over long distances. Such a quantum internet will be useful for
distributed quantum computing, distributed quantum cryptographic protocols and dramatically
lowering communication complexity. Since quantum information processing can offer exponential
increase in computing power for some tasks as well as unconditional security, the realization of a
quantum internet is an important long-term scientific and technological goal. Thanks to the long
coherence time of photons, photonic channels, for example, optical fibers are often used for quantum
communication. Nonetheless, owing to losses, the probability of successful transmission of a photon
through an optical fiber decays exponentially. 
Consequently, the efficiency of this kind of quantum communication decreases exponentially with the communication distance, which is
limited to hundreds of kilometers \cite{SSRG09}.

\begin{figure}[b]  
\includegraphics[keepaspectratio=true,height=30mm]{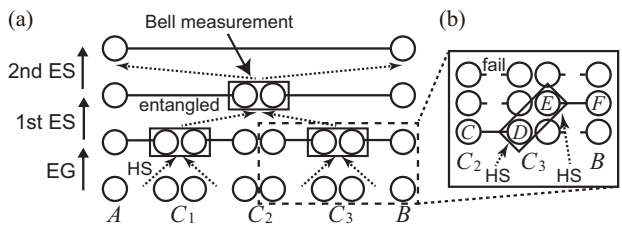}
  \caption{(a) Conventional protocols \cite{B98,DLCZ,SSRG09,C06,K08,ATKI10,ZDB12,M10,L12}, and (b) parallel preparation of entangled pairs. (a) The repeater protocol is a way to supply entanglement with two-end parties, Alice ($A$) and Bob ($B$), via repeater nodes $\{ C_i \}_{i=1,2,\ldots,n}$ ($n=3$ here).
The protocol starts with entanglement generation (EG) through transmitting photons between adjacent repeater nodes, 
followed by recursive applications of the entanglement swapping (ES).
The ES at a repeater node (e.g., 2nd-round ES) starts only after
the node receives signals for heralding the successful preparation of two entangled pairs at the previous round (1st-round ES).
This preparation may be executed in a parallel fashion as in (b) by using multiple quantum memories,
where the heralding signals (HSs) are used to pick up an appropriate pair.
Even in this way,
if the EG and ES succeed only probabilistically as in Refs.~\cite{DLCZ,SSRG09,K08,ATKI10},
the total classical communication time for the transmission of the HSs alone is proportional to the total distance $L$.
Since the matter qubits generally decay exponentially with this time,
the protocols \cite{DLCZ,SSRG09,K08,ATKI10} scale exponentially with (the square root of \cite{R09}) the distance $L$ between Alice and Bob (irrespectively of the employed purification schemes \cite{R09}).
If HSs are not exchanged, the protocol \cite{B98,DLCZ,SSRG09,C06,K08,ATKI10,ZDB12,M10,L12} is merely the quantum relay \cite{W02} with exponential scaling.
}
  \label{fig:1.eps}
\end{figure}

To overcome such a distance limit, quantum repeaters that use repeater nodes between the sender (Alice) and the receiver (Bob) are needed \cite{B98} in order to enjoy the blessing, i.e., the polynomial scaling of the efficiency with the total distance. In contrast to conventional repeaters in the classical communication, quantum repeaters cannot make redundant copies of quantum signals due to the no-cloning theorem \cite{WZ82}.
Instead, as shown in Fig.~\ref{fig:1.eps}(a), the standard approach \cite{B98,DLCZ,SSRG09,C06,S09,K08,ATKI10,ZDB12,M10,L12,J09} to quantum repeaters equips the repeater nodes with quantum memories,
and starts with entanglement generation for the quantum memories between adjacent nodes via the transmission of {\it photons entangled with the memories}.
Then, entanglement swapping \cite{Z93} is, one after another, performed at a node that has confirmed the existence of entanglement with other repeater nodes {\it by receiving heralding signals from different repeater nodes at long distances}.
Thus, the quantum memories are at least required (i) to be entangled with photons (perhaps with a {\it telecom} wavelength for the fiber transmission) for the entanglement generation, and (ii) to be able to {\it preserve entanglement faithfully} at least until receiving the heralding signals for the entanglement swapping from the {\it distant} nodes.
Note that, without such quantum memories, the repeater protocols are, at least, inevitably reduced into quantum relay protocols \cite{W02} with the exponential scaling (see Fig.~\ref{fig:1.eps}).

Earlier proposals \cite{DLCZ,SSRG09} (including the seminal paper \cite{DLCZ} of Duan {\it et al.}) for the realization regard an atomic ensemble as such a quantum memory with {\it infinite} coherence time for (ii), relying on a {\it probabilistic} Bell measurement on single photons that are entangled with the memories via collectively enhanced coupling \cite{DLCZ} for (i).
The protocols \cite{DLCZ,SSRG09} have a beauty of
simplicity in terms of the numbers of repeater nodes and the required matter quantum memories. 
However, unfortunately, if the coherence time of the matter quantum memories is {\it finite}, 
those simple protocols \cite{DLCZ,SSRG09} are shown \cite{R09} to scale {\it exponentially} with (the square root of) the communication distance (irrespectively of employed purification schemes \cite{R09}).
This is caused by the necessity of the transmission of heralding signals whose communication time is proportional to {\it the total distance} due to the probabilistic nature of the Bell measurements \cite{DLCZ,SSRG09}, and by the {\it exponential} dephasing or depolarization on the matter quantum memory with time (see Fig.~\ref{fig:1.eps}).
Thus, only remaining solutions to overcome the problem in those simple protocols \cite{DLCZ,SSRG09} would be (I) to boost the success probability of the Bell measurement (e.g., to invoke a near-deterministic Bell measurement \cite{KLM01} on single photons) or (II) to make the coherence time infinite by equipping the matter quantum memory with fault tolerance. 
But, either of these spoils the claimed simplicity of the original proposals \cite{DLCZ,SSRG09}.

A simple solution for (I) or (II) may be to use matter qubits satisfying Divincenzo's 2nd-to-5th criteria \cite{D00} (initialization, quantum gates faster than decoherence time, universal gate set, and readout) as the matter quantum memories,
as in the protocols \cite{B98,C06,S09,ATKI10,ZDB12,M10,M12,L12,G12,J09}.
In fact, some \cite{J09,M10,M12,G12} of them have already been shown to work even with finite-coherence-time matter qubits.
However, unfortunately,
the matter qubits are normally less efficient \cite{DLCZ} in the coupling with photons for (i) than the atomic ensembles, and the efficient coupling remains a very challenging technology {\it even with atomic ensembles} in spite of recent experimental advances \cite{L10,K08,S10,MK13}.
Thus, we have not yet been able to refute Divincenzo's conjecture \cite{D00} that the efficient coupling between a matter qubit and photons for (i)---corresponding to Divincenzo's {\it extra} criterion \cite{D00,L10}---is really hard.
The only theoretical solution \cite{M10} to compensate this inefficiency to entanglement generation under reasonable coherence time is to use a lot of matter qubits at each repeater node like the protocols \cite{M10,M12,G12,J09}, i.e., to require the matter qubits to satisfy even Dinvincenzo's 1st criterion (scalability). 
However this implies that the matter qubits in the quantum repeaters \cite{B98,C06,S09,ATKI10,ZDB12,M10,M12,L12,G12,J09} need to satisfy not only Divincenzo's five (1st-5th) criteria \cite{D00} for universal quantum computation but also his (really hard) extra criterion.
Therefore, the quantum repeaters \cite{B98,C06,S09,ATKI10,ZDB12,M10,M12,L12,G12,J09} could be more difficult than universal quantum computation.
This is the statement coming from the dogma \cite{B98,DLCZ,SSRG09,C06,S09,ATKI10,ZDB12,M10,M12,L12,G12,J09} of the requirements of matter quantum memories for quantum repeaters,
which will remain undeniable without a future experimental breakthrough.

The main point of this paper is to disprove such a dogma that a demanding matter quantum memory is necessary for accomplishing quantum repeaters.
Here, we present a scheme which shows that it is possible to achieve an all photonic quantum repeater with flying qubits only. 
Our scheme uses only single-photon sources, linear optical elements, photon detectors, and a fast active feedforward technique (less than 150 ns \cite{P07}), similarly to optical universal quantum computation \cite{KLM01,VBR06}.
However, we provide evidence that our all photonic quantum repeaters are easier than universal quantum computation \cite{KLM01,VBR06}, in contrast to the conventional quantum repeaters \cite{B98,DLCZ,SSRG09,C06,S09,K08,ATKI10,ZDB12,M10,M12,L12,G12,J09}.
In addition, the all photonic feature of our repeaters has the following advantages that can never be obtained in the quantum repeaters based on matter quantum memories \cite{B98,DLCZ,SSRG09,C06,S09,K08,ATKI10,ZDB12,M10,M12,L12,G12,J09}: 
(a) The heralding signals for achieving the entanglement swapping are sent and received within the {\it same} repeater nodes, rather than between different repeater nodes at long distances, which reduces the transmission distance and time of the heralding signals to zero in principle.
(b) Combined with a machine-gun-like \cite{L09} single-photon source, this feature allows us to increase the repetition rate of our protocol as high as one wants.
(c) Even if we use a single-photon source based on a matter qubit, 
the matter qubit is no longer required to have a deterministic interaction with photons as well as to
have long coherence time (and, of course, a matter {\it quantum memory} \cite{S10,SSRG09}  can be diverted to a single-photon source), let alone to satisfy Divincenzo's all the criteria \cite{D00}.
(d) Frequency converters for photons to increase the coupling to matter quantum memories \cite{T05} and to optical fibers \cite{I11} could be unnecessary.
(e) Our protocol could work at room temperature.

We draw our protocol from a concept, ``time reversal,'' underlying the distinguished findings in quantum information theory, such as the measurement-based quantum computation \cite{GC99,RB01} and the measurement-device-independent quantum key distribution (QKD) \cite{LCQ12}.
In fact, 
our protocol corresponds to the time reversal of the conventional quantum repeaters \cite{B98,DLCZ,SSRG09,C06,S09,K08,ATKI10,ZDB12,M10,L12,G12,J09}, where entanglement swapping is performed before entanglement generation. 
This is an innovative part of our proposal.
As an example to achieve such a time-reversed quantum repeater, we use cluster-state \cite{RB01} flying qubits rather than simple Bell pairs, in contrast to existing quantum repeaters \cite{B98,DLCZ,SSRG09,C06,S09,K08,ATKI10,ZDB12,M10,L12,J09}. 
Since our protocol is the time-reversed version of a conventional quantum repeater with a polynomial scaling,
our protocol follows the same scaling.

\section{Conventional Quantum Repeaters}
We start by considering the essential of the polynomial scaling of the conventional quantum repeaters (see Fig.~\ref{fig:1.eps}), i.e., 
the execution of the entanglement swapping {\it upon confirming} the existence of entangled pairs.
Entanglement swapping is a way to share an entangled pair over a longer distance through connecting two (short) entangled pairs.
Given an entangled state between
systems $C$ and $D$ and an entangled state between systems $E$ and $F$ (Fig.~\ref{fig:1.eps}(b)), it is possible to establish
entanglement between systems $C$ and $F$, by performing the Bell measurement on the systems $D$ and $E$.
Hence, if distances between $CD$ and between $EF$ are $l$ and if $DE$ is held at a single node, 
the entanglement swapping presents an entangled pair $CF$ separated by distance $2l$.

If we regard this entanglement swapping as the one implemented in a round of a quantum repeater protocol (Fig.~\ref{fig:1.eps}(a)), 
the entangled pairs $CD$ and $EF$ correspond to those prepared through the success of all the relevant entanglement generations and all the previous rounds of entanglement swapping.
These entanglement preparations can be {\it repeatedly} applied to the {\it specific} qubits $CD$ and $EF$ until they are successfully entangled, as in proposed protocols \cite{DLCZ,SSRG09,ATKI10}.
However, in this case, owing to the fact that the entanglement preparations for $CD$ and for $EF$ are independent and merely probabilistic processes,
the timings of successfully producing the entangled pairs $CD$ and $EF$ are not necessarily the same, which would require additional memory time for waiting the joint success event.

Instead, we can use a parallel procedure (as in Fig.~\ref{fig:1.eps}(b)) to synchronize the successes of the entanglement preparations.
In this method, each of the entanglement preparations for $CD$ and for $EF$ is executed {\it in parallel} by applying it to a sufficiently large number of qubits in order to successfully produce at least one entangled pair.
Then, the prepared entangled pairs to be referred to as $CD$ and $EF$ appear {\it simultaneously}.
Although this method reduces the requirements for the memory time of qubits,
it still requires the qubits to have long memory time. 
In fact, the node to perform the Bell measurement on the counterparts $DE$ needs to wait for the arrivals of
heralding signals for specifying the qubits $DE$ among many candidates at the same node (as shown in Fig.~\ref{fig:1.eps}(b)).
Then, as inferred by Fig.~\ref{fig:1.eps}(a), we notice an inherent problem of the quantum communication:
The heralding signals should travel over long distances. 
This transmission time
is at least the classical communication time between adjacent repeater nodes, and can be extended
to the order of the communication time over the total distance if the entanglement swapping works
only probabilistically as in simple schemes \cite{DLCZ,SSRG09,K08,ATKI10}.
Due to this waiting time, the conventional quantum repeaters \cite{B98,DLCZ,SSRG09,C06,S09,K08,ATKI10,ZDB12,M10,L12,G12,J09} need memory time and the repetition rate is limited.

\begin{figure}[b] 
\includegraphics[keepaspectratio=true,height=35mm]{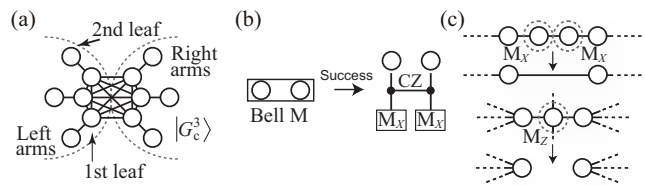}
  \caption{(a) Complete-like cluster states $\ket{G_{\rm c}^m}$ (for the case of $m=3$). The state $\ket{G_{\rm c}^m}$ has $2m$ arms, each of which is composed of 1st-leaf and 2nd-leaf qubits. 
The edge represents the past application of the controlled-$Z$ (CZ) gate to qubits initialized in state $(\ket{H}+\ket{V})/\sqrt{2}$, implying the existence of entanglement between them \cite{RB01}. Here, $\ket{H}$ and $\ket{V}$ represent a basis of a single-photon qubit.
The 1st-leaf qubits correspond to the memories held by a single repeater node in the conventional repeaters (e.g., repeater node $C_3$ in Fig.~\ref{fig:1.eps}(b)).
Single photons belonging to left arms (right arms) are to be sent to the left-hand-side (right-hand-side) adjacent receiver node (see Fig.~\ref{fig:3.eps}).
  (b) Bell measurement base
d on linear optical elements and photon detectors \cite{W94}. If it succeeds, it works as the CZ gate followed by the $X$-basis measurements. If it fails, $Y$-basis measurements are applied for the existing photons, and, for lost photons, it informs us of the photon losses.  
  (c) Two adjacent $X$-basis measurements M$_X$ on a linear cluster remove the
qubits and directly connect their neighbors \cite{RB01}. The $Z$-basis measurement M$_Z$ on a qubit removes the qubit \cite{RB01}.
}
  \label{fig:2.eps}
\end{figure}

The long waiting time for the transmission of the heralding signals becomes a problem even for an all photonic quantum repeater because the waiting time corresponds to the losses for the photonic qubits.
To overcome this problem, we introduce a time-reversed all-photonic version of the conventional quantum repeaters \cite{DLCZ,SSRG09,K08,ATKI10},
where the waiting time could be made zero in principle.

\section{Time-Reversed All-Photonic Quantum Repeaters}
Let us begin with specifying the role of the Bell measurement on the counterparts $DE$ of the entangled pairs $CD$ and $EF$ in the parallel procedure of Fig.~\ref{fig:1.eps}(b).
Here, the Bell measurement implicitly plays a role to entangle qubits $D$ and $E$ at a moment, as it can be regarded as an entangling operation followed by $X$-basis measurements (see Fig.~\ref{fig:2.eps}(b)).
Then, the time reversal of the whole process may be as follows:
we first generate entanglement between $DE$, and then create entanglement between $CD$ and between $EF$, which is followed by $X$-basis measurements on $DE$.
However, at the beginning of this time-reversed protocol, 
it is impossible to specify the qubits $DE$ among the many candidate qubits at the same node (as shown in Fig.~\ref{fig:2.eps}(b)), 
because the heralding signals for the specification will be given after the successful entanglement preparations between $CD$ and between $EF$.
Thus, we propose to use the cluster state $\ket{G_{\rm c}^m}$ that has $2m$ arms composed of 1st-leaf and 2nd-leaf qubits (see Fig.~\ref{fig:2.eps}(a)). Here the 1st-leaf qubits serve as the candidate qubits at the same node 
and any pair of the 1st-leaf qubits is completely connected by edges that respectively show the existence of entanglement. 
Then, since every pair of the 1st-leaf (candidate) qubits in the state $\ket{G_{\rm c}^m}$ is already entangled, 
in contrast to conventional repeater protocols \cite{B98,DLCZ,SSRG09,C06,S09,K08,ATKI10,ZDB12,M10,L12,G12,J09},
we are not required to perform the (possibly probabilistic) Bell measurement on the qubits $DE$, let alone to specify the qubits $DE$ {\it in advance}.
Thus, the only remaining task at this point is to execute the $X$-basis measurements on the qubits $DE$ according to the heralding signals that are to be given {\it later}.

\begin{figure*}[t]  
\includegraphics[keepaspectratio=true,height=25mm]{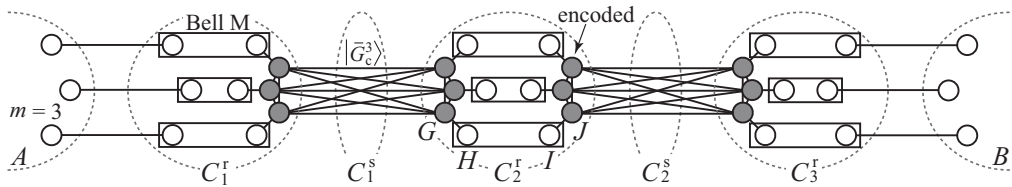}
  \caption{Snapshot of all photonic quantum repeaters [$(m,n)=(3,2)$].
The protocol is defined as follows: 
(i) Alice (Bob) prepares $m$ single photons that are maximally entangled with her (his) local qubits and sends them to the adjacent receiver node $C_1^{\rm r}$ ($C_{n+1}^{\rm r}$). 
At the same time, any other source node  $C^{\rm s}_i$ prepares the encoded complete-like cluster state $\ket{\bar{G}_{\rm c}^m}$, 
and the left (right) arms are sent to the left-hand (right-hand) adjacent receiver node $C_i^{\rm r}$ ($C_{i+1}^{\rm r}$).
(ii) On receiving the single photons (this moment is snapshotted), every receiver node applies the Bell measurement of Fig.~\ref{fig:2.eps}(b) on the $m$ pairs of the 2nd-leaf qubits of the left and right arms.
(iii) If one of the Bell measurements succeeds, the receiver node performs the loss-tolerant $X$-basis measurements on the 1st-leaf qubits on the successful arms, 
and makes the loss-tolerant $Z$-basis measurements on all the other 1st-leaf qubits. If all the $m$ Bell measurements or one of the loss-tolerant measurements on the 1st-leaf qubits fails, the receiver node considers that this trial fails.
(iv) Finally, the receiver nodes announce all the measurement outcomes to Alice and Bob, and the protocol succeeds when no receiver node judges this trial as failure.
}
  \label{fig:3.eps}
\end{figure*}

As we have seen, the 1st-leaf qubits of the state $\ket{G_{\rm c}^m}$ correspond to quantum memories at a single repeater node in the conventional repeaters.
In this analogy, the 2nd-leaf qubits serve as the single photons to supply entanglement to the 1st-leaf qubits between adjacent repeater nodes, i.e., they are used for entanglement generation process. 
To see this, let us consider a process to connect 1st-leaf qubits $G$ and $J$ in repeater node $C_2^{\rm r}$ of Fig.~\ref{fig:3.eps}.
Since the 1st-leaf qubits $G$ and $J$ are respectively entangled with the 2nd-leaf qubits $H$ and $I$, 
if a linear-optics-based Bell measurement of Fig.~\ref{fig:2.eps}(b) on the 2nd-leaf qubits $H$ and $I$ succeeds,
the 1st-leaf qubits $G$ and $J$ are entangled,
and they become the candidates for the qubits $DE$ that are to receive the $X$-basis measurements.
On the other hand, if the Bell measurement fails owing to the photon losses of the 2nd-leaf qubits or the bunching effect of the photons,
we apply $Z$-basis measurements on the 1st-leaf qubits $GJ$.
This $Z$-basis measurements remove the corresponding arms without affecting the entanglement structure of the other arms of $\ket{G_{\rm c}^m}^{\otimes 2}$ according to the rule of Fig.~\ref{fig:2.eps}(c). 
This connection process for the 1st-leaf qubits can be executed in parallel for any arm of the state $\ket{G_{\rm c}^m}$, which corresponds to
the parallel entanglement generation (Fig.~\ref{fig:1.eps}(b)) in the conventional quantum repeaters in Fig.~\ref{fig:1.eps}(a).

Note that the connection process requires the heralding signals from the 2nd-leaf qubits to the 1st-leaf qubits.
If the 1st-leaf qubits were matter qubits that are stationary at a repeater node,
the heralding signals would still be exchanged between adjacent repeater nodes, requiring 
the transmission time ranging from hundred microseconds to milliseconds.
Thus, the role of the 1st-leaf qubits could be still challenging for matter-qubit quantum memories from the current status \cite{S10}.
However, in our proposal, the 1st-leaf qubits are composed of single-photon qubits.
Thus, the 1st-leaf qubits can be sent with the 2nd-leaf qubits, 
which holds the transmission time of the heralding signals to a minimum, i.e., the {\em local} active feedforward time. 
However, this causes an alternative problem that we need to apply single-qubit measurements on the 
the 1st-leaf qubits {\it faithfully} even under the {\it photon losses} as well as {\it small errors}
of the transmission.
But, since the transmission is performed merely between adjacent repeater nodes and the losses and the channel errors are thus independent of the total distance, 
they can be overcome by invoking a loss-tolerant scheme to execute a single-qubit measurement for the 1st-leaf qubits, 
say a protocol of Varnava {\em et al.} \cite{VBR06}.
More specifically, instead of the state $\ket{G_{\rm c}^m}$, we use its encoded version  $\ket{\bar{G}_{\rm c}^m}$, i.e., the complete-like cluster state $\ket{\bar{G}_{\rm c}^m}$ with the encoded 1st-leaf qubits that are colored in gray in Fig.~\ref{fig:3.eps}. 
An explicit method to prepare the state $\ket{\bar{G}_{\rm c}^m}$ locally with linear optical elements, single-photon sources, photon detectors, and a high-speed active feedforward technique is given in Secs.~\ref{se:basic} and \ref{se:preparation} of Appendix, which determines the preparation time that is to be translated into the corresponding photon-loss probability.

The encoding of the protocol of Varnava {\em et al.} \cite{VBR06} is done by replacing a qubit being to receive a single-qubit measurement under loss with an encoded qubit composed of plural physical qubits.
The loss-tolerant measurement is performed with an arbitrary high success probability via only single-qubit measurements on the physical qubits, as long as the loss probability for the physical qubits is less than 50 \% (corresponding to the loss of a 15-km optical fiber).
Thus, in our protocol, the loss for the 1st-leaf qubits should be less than 50 \% 
by adjusting the transmission distance. 
This limitation corresponds to an analogy of the one on the quantum memory in the conventional quantum repeaters, 
although they differ in the types of noises [loss and depolarization (or dephasing)].

In addition to the tolerance to the loss, as seen in
Secs.~\ref{se:error-rud} and \ref{se:num-rud} of Appendix,
remarkably, it turns out that the scheme of Varnava {\it et al.} \cite{VBR06}
allows us to perform $Z$-basis or
$X$-basis measurement {\it faithfully even under general errors}.
Thus, this scheme highly fits with our repeater scheme that needs
(loss-tolerant) $Z$-basis or $X$-basis measurements only.

Notice, however, that Varnava {\it et al.}'s scheme is less
robust and loss-tolerant when non-Pauli measurements are performed. Since
universal optical quantum computing \cite{KLM01} requires such
non-Pauli measurements, Varnava {\it et al.}'s scheme
requires more overhead and has a much lower error threshold
in the case of universal optical quantum computing.
This highlights the difference in the performance
of Varnava {\it et al.}'s scheme in the two applications---quantum repeaters and universal optical quantum computation.

To see how our protocol runs more precisely, 
we describe the whole protocol.
In the repeater, all the repeater nodes between Alice and Bob separated by distance $L$ are classified into two sets, called source nodes $\{ C^{\rm s}_i\}_{i=1,\ldots,n}$ and
receiver nodes $\{ C^{\rm r}_i\}_{i=1,\ldots,n+1}$.
The source nodes and the receiver nodes are placed alternatively and at regular intervals, and adjacent source nodes (adjacent receiver nodes) are separated, say $L_0=L/(n+1)$ apart.
In addition, the arms of the state $\ket{G_{\rm c}^m}$ are classified into the right-hand and left-hand sets as in Fig.~\ref{fig:2.eps}(a). 
Then, the repeater runs according to the protocol in Fig.~\ref{fig:3.eps}.

\section{Applications}
If we use our protocol in Fig.~\ref{fig:3.eps} for QKD,
since Alice and Bob's raw key is virtually regarded as the one that is obtained by measurements on Alice and Bob's entangled pairs {\it before} starting our repeater protocol, Alice and Bob's qubits in Fig.~\ref{fig:3.eps} can be virtual \cite{LCQ12}. 
Moreover, since a possible correction to their qubits after the protocol is merely the application of a (unitary) Pauli operation, this correction corresponds to bit flips on their raw key.
In addition, since every repeater node requires no classical communication with the other nodes according to the protocol of Fig.~\ref{fig:3.eps},
the time required for each trial of the protocol is determined only by the number of the local active feedfowards used in steps (ii) and (iii) of the protocol.
But this is merely one time (of the Bell measurements), because the loss-tolerant $X$-basis and $Z$-basis measurements in the step (iii) require no active feedforward (see Sec.~\ref{se:error-rud} of  Appendix).

Without any need of quantum memories, our all photonic quantum repeater scheme works not only in QKD, but also in many other quantum information processing protocols such as non-local measurements \cite{V03} and cheating strategies \cite{LL11} in position-based quantum cryptography \cite{B11}. In those protocols, entanglement, once generated, is consumed immediately to generate classical output strings. For this reason, no quantum memory is needed in the protocol (see a flexibility of our repeater protocol in Sec.~\ref{se:tot} of Appendix). Furthermore, Pauli errors can be taken care of off-line (i.e., in the classical communication phase of the protocol).

For protocols that demand strictly a quantum output state, of course, quantum memories are needed. For instance, suppose Alice would like to transfer a quantum state to a distant observer, Bob via quantum teleportation \cite{B93}. Suppose further that Bob insists on keeping the final state as a quantum state (as he has no idea what measurement, if any, he might wish to perform in future). In this case, the very fact that the final state is quantum means that the protocol requires effectively quantum memories with memory time in the order of classical communication time between Alice and Bob.
However, even in this case,
in contrast to the standard quantum repeaters \cite{B98,DLCZ,SSRG09,C06,S09,K08,ATKI10,ZDB12,M10,L12,J09} as in Fig.~\ref{fig:1.eps},
the memory time required in the quantum teloportation based on our repeater protocol scales only {\it linearly} with communication distance $L$ like the speediest protocol \cite{M12}, differently from {\it polynomial} or {\it subexponential} scalings of the conventional ones \cite{DLCZ,SSRG09,ATKI10,C06},
which leads to greater suppression of the errors of the quantum memories (see the details in Sec.~\ref{se:tot} of Appendix).

\section{Scaling and Performance}
As expected from the time-reversed-like construction of our protocol itself, 
the average of the total photon number $\bar{Q}$ consumed in our protocol to produce an entangled pair between Alice and Bob scales only polynomially with the total distance. 
In addition, the average rate $\bar{R}$ of our protocol to produce an entangled pair with a single repeater system is in the order of the repetition rate $f$ of the slowest devices among single-photon sources, photon detectors, and active-feedforward techniques,
which is in a striking contrast to the conventional repeaters \cite{B98,DLCZ,SSRG09,C06,S09,K08,ATKI10,ZDB12,M10,L12} whose rates are restricted by the communication time between adjacent nodes (from hundred microseconds to milliseconds), at least.
Moreover, individual depolarization caused by the channel errors is suppressed exponentially, thanks to the special robustness of the protocols of Varnava {\it et al.} for $Z$-basis and $X$-basis measurements and to faithful transmission of a single photon over a short distance.
The details are given in Sec.~\ref{se:repeater} of Appendix.

To show the scaling of our protocol explicitly, we present $\bar{Q}$, $\bar{R}$, and the average fidelity $\bar{F}$ of the obtained entangled pair for two cases. Here we assume that photons always run in optical fibers with the transmittance $T=e^{-l/l_{\rm att}}$ for distance $l$ ($l_{\rm att}=22$ km).
In addition, we suppose that the optical fibers have small errors when they are used to connect distant repeater stations ($L_0/2$ apart), and the errors of the fiber with length $L_0/2$ can be described as an individual depolarizing channel with error probability $e_{\rm d}$.
We also assume to use single photon sources with efficiency $\eta_{\rm S}$, photon detectors with quantum efficiency $\eta_{\rm D}$, and active feedforward techniques less than 150 ns.
For the choice of $L=5000$~km ($L=1000$~km), $L_0=4$~km, $e_{\rm d}=4.2 \times 10^{-5}$,
$\eta_{\rm D} \eta_{\rm S}=0.95$, $f=100$~kHz, and $m=24
$ ($m=19$), we obtain $\bar{Q}=4.0 \times 10^7$ ($\bar{Q}=4.1 \times 10^6$),
$\bar{R}=69$~kHz ($\bar{R}=58$~kHz), and $\bar{F}=0.90$ ($\bar{F}=0.97$) under numerical calculation to minimize $\bar{Q}$
(see the detail in Sec.~\ref{se:num} of Appendix).
On the other hand, if Alice and Bob use the direct transmission of single photons emitted by a 10~GHz single-photon source, 
in order to share an entangled pair,
they need to consume, on average, $5.1 \times 10^{98}$ ($5.5 \times 10^{19}$) single photons 
and to take $10^{81}$ ($175$) years. 
This striking contrast highlights the exponential superiority of our repeater protocol to the existing photonic protocols \cite{W02}.
In addition, the rate $\bar{R}$ of our protocol is at least 5 order of magnitude better than those of the standard repeater schemes \cite{DLCZ,SSRG09,ATKI10,C06,S09}.
Moreover,
our protocol is comparable to the speediest protocol of Munro {\it et al.} \cite{M12} in the rate  $\bar{R}$, 
although the protocol \cite{M12} uses not only single photons but also demanding matter qubits and their required numbers are in the same order of the consumed photons $\bar{Q}$ in our protocol (see Sec.~\ref{se:tot} of Appendix).

\section{Discussion}
So far, the requisites for existing quantum repeater protocols---such as infinite coherence time for matter quantum memories \cite{DLCZ,SSRG09}, an on-demand emission of a single photon from the matter quantum memory \cite{DLCZ,SSRG09,C06,S09}, the use of a single-photon source \cite{SSRG09,M12}, 
Divincenzo's {\it all} the criteria \cite{D00} beyond his five criteria for universal quantum computation \cite{B98,C06,S09,K08,ATKI10,ZDB12,M10,L12,M12,G12,J09},
and a reversible quantum interface \cite{T05,I11} between photons with different wave lengths \cite{B98,DLCZ,SSRG09,C06,K08,ATKI10,ZDB12,M10,L12,M12,G12,S09,J09}---may have been too many to be satisfied. 
Since some of those techniques could directly lead to the realization of a single-photon source that is just needed in our repeater protocol (e.g., a matter quantum memory can be used as a generator of hundreds \cite{SSRG09} or even thousands \cite{L09} of single photons), the developments would be significant even for our scheme. 
However, our protocol greatly and certainly reduces the number of requirements for quantum repeaters.
Even from a fundamental viewpoint, 
the all photonic feature of our theory enables single photons to fully describe even quantum repeaters
in addition to quantum computation \cite{KLM01} and boson sampling \cite{B13,S13},
which leads to the first rigorous proof  that a quantum repeater is much simpler than a quantum computer.
We have only just begun to grasp the full implications of all photonic quantum repeaters: e.g., 
a proposal for a good single-photon source,
a proof-of-principle experiment with photons with a telecom wavelength,
a more experimentally oriented modification of our protocol, 
a more robust improvement against noises, and a generalization for a general network topology including a two dimensional lattice or an irregular two-dimensional graph will lead to an attractive new twist.

\section*{Acknowledgements}
We thank K.~Fujii, D.~Gottesman, R.~Jozsa, G.~Kato, M.~Koashi, M.~Owari, K.~Shimizu, H.~Takesue, and S.~Tani for valuable discussions, and especially M.~Curty and W.~J.~Munro for fruitful discussions in the early stage of this study.
This research is in part executed under the Project UQCC by the National Institute of Information and Communications Technology (NICT).
K.A. is supported in part by a MEXT Grant-in-Aid for Scientific Research on Innovative Areas 21102008, and
K.T. is supported in part by the Japan Society for the Promotion of Science (JSPS) through its Funding Program  for World-Leading Innovative R\&D on Science and Technology (FIRST Program). 
H.-K.L. thanks financial support from NSERC and CRC program.

\appendix

\vspace{8mm}
\begin{center}
{\bf Appendix}
\end{center}
\vspace{3.3mm}

Here we present the detailed analysis of our all photonic quantum repeater protocol defined in Fig.~\ref{fig:3.eps}. The protocol is based on the local preparation of the encoded completely-like cluster state $\ket{\bar{G}_{\rm c}^m}$ at the source nodes.
As defined in Figs.~\ref{fig:2.eps}(a) and \ref{fig:3.eps}, the state $\ket{\bar{G}_{\rm c}^m}$ has $2m$ arms, each of which is composed of an {\it encoded} 1st-leaf qubit and a 2nd-leaf qubit.
The encoding for the 1st-leaf qubits follows the protocol of Varnava {\em et al.} \cite{VBR06},
which allows us to perform $Z$-basis and $X$-basis measurements on them {\it almost deterministically} and {\it faithfully} even under loss and general errors.
The fact that our protocol uses only these $Z$-basis and $X$-basis measurements with this {\it special robustness} implies that our protocol is much easier than universal quantum computation requiring general single-qubit measurements.
We show these facts in Sec.~\ref{se:var} via reconsidering the protocol of Varnava {\em et al.} \cite{VBR06}.
In Sec.~\ref{se:preparation}, we provide an explicit method to prepare the encoded completely-like cluster state $\ket{\bar{G}_{\rm c}^m}$ locally with linear optical elements, single-photon sources, photon detectors, and a high-speed active feedforward technique.
This method is basically the same as the protocol of Varnava {\em et al.} \cite{VBR08}.
In Sec.~\ref{se:repeater}, using the results in Secs.~\ref{se:var} and \ref{se:preparation},
we present the detailed analysis of our all photonic quantum repeater protocol for determining its performance, as well as a comparison with the speediest protocol given by Munro {\it et al.} \cite{M12}.

\section{Loss-tolerant single-qubit measurement}\label{se:var}

In our repeater protocol defined in Fig.~\ref{fig:3.eps}, we use the protocol of Varnava {\it et al.} \cite{VBR06} to perform $Z$-basis and $X$-basis measurements on the 1st-leaf qubits of the encoded completely-like cluster state $\ket{\bar{G}_{\rm c}^m}$.
This protocol has originally been proposed to execute a single-qubit measurement of an observable $\hat{A}(\alpha)=\cos \alpha \hat{X}+\sin \alpha \hat{Y}$ on a qubit in a cluster state with an arbitrary high success probability under the loss, and it {\it does not} have robustness against general errors in general.
However, if this protocol is used to implement $Z$-basis or $X$-basis measurement,
the protocol can be so equipped with a majority vote as to have robustness against general errors, as briefly mentioned by Varnava {\it et al.} \cite{VBR06}.
In this section, we prove this special robustness of Varnava {\it et al.}'s protocol \cite{VBR06}, through reviewing basic results of Ref.~\cite{VBR06} in Sec.~\ref{se:basic},
elaborating the analysis of the effects of general errors on the protocol equipped with the majority vote in Sec.~\ref{se:error-rud}, 
and then showing the numerical examples in Sec.~\ref{se:num-rud}.

\subsection{Basic results \cite{VBR06}}\label{se:basic}

\begin{figure}[b]
\includegraphics[keepaspectratio=true,height=28mm]{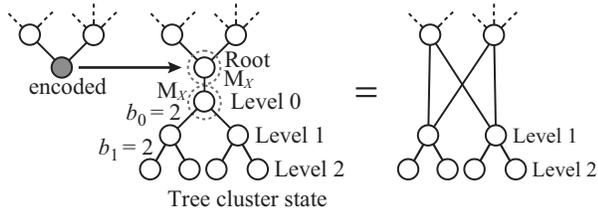}
\caption{Example of the encoding of a qubit on a tree cluster state with branching parameters $b_0=b_1=2$. The gray qubit corresponds to the encoded qubit. 
$X$-basis measurements on the root and 0th-level qubits should be applied in advance to complete the encoding, linking every 1st-level qubit to all the qubits that have been connected to the encoded qubit.}
  \label{fig:a2.eps}
\end{figure}

Let us begin by briefly reviewing the basic results shown in Ref.~\cite{VBR06}.
The loss-tolerant measurement \cite{VBR06} is a way to execute the single-qubit measurement of an observable $\hat{A}(\alpha)=\cos \alpha \hat{X}+\sin \alpha \hat{Y}$ on a qubit in a cluster state under the loss.
The qubit to be equipped with the function of the loss-tolerant measurement, called encoded qubit, should be replaced by the root qubit of a tree cluster state with branching parameters $\{b_i\}_{i=0,1,\ldots,l}$ like Fig.~\ref{fig:a2.eps}, 
and, further, $X$-basis measurements on the root and 0th-level qubits should be applied in advance.
These $X$-basis measurements connect every 1st-level qubit to all the qubits that have been linked to the encoded qubit.
The number $Q_{\rm L}$ of qubits in the tree cluster state
is 
\begin{equation}
Q_{\rm L}= \sum_{j=0}^l \prod_{i=0}^j b_i. \label{eq:var-1}
\end{equation}
The loss-tolerant measurement is executed by applying proper single-qubit measurements on the qubits below the 1st level in the tree. 
According to Ref.~\cite{VBR06}, for (individual) loss probability $\epsilon_0$, the success probability $P_{\rm L}$ of the loss-tolerant measurement of observable $\hat{A}(\alpha)$ is described by
\begin{equation}
P_{\rm L}=[(1-\epsilon_0 +\epsilon_0 R_1)^{b_0} - (\epsilon_0 R_1)^{b_0}] (1-\epsilon_0+\epsilon_0 R_2)^{b_1}, \label{eq:PL}
\end{equation}
where $R_k$ is the success probability of implementing an indirect $Z$-basis measurement on any given qubit found in the $k$th level of the tree, specified by
\begin{equation}
R_k = 1- [1-(1-\epsilon_0 ) (1-\epsilon_0 +\epsilon_0 R_{k+2})^{b_{k+1}}]^{b_k}
\label{eq:rk}
\end{equation}
through $1 \le k\le l$, $R_{l+1}:=0$, and $b_{l+1}:=0$.
Note that $P_{\rm L}$ of  Eq.~(\ref{eq:PL}) can be determined by solving the equations of (\ref{eq:rk}) recursively.
The success probability $P_{\rm L}$ of this measurement can be made arbitrary close to unity, as long as $\epsilon_0<0.5$, 
and it is numerically shown \cite{VBR06,VBR07} to be
\begin{equation}
Q_{\rm L} \simeq {\rm poly} \ln \frac{1}{1-P_{\rm L}} \simeq \left( \ln \frac{1}{1-P_{\rm L}}\right)^{4.5}.  \label{eq:QI}
\end{equation}

Note that, for a qubit in the tree cluster state, we can perform $Z$-basis measurement even if we lose the qubit. This {\it indirect $Z$-basis measurement} is essential for the protocol of Varnava {\em et al.} \cite{VBR06}.
Let $N_a$ be the set of qubits that are (directly) connected to qubit $a$.
The indirect  $Z$-basis measurement on a qubit $A$ in the $k$th level of the tree is achieved by the joint success of direct $X$-basis measurement on {\it any} qubit $B \in N_A$ in the $(k+1)$th level and direct or indirect $Z$-basis measurements on $b_{k+1}$ qubits $\{C^B_i\}_{i=1.\ldots b_{k+1}}$ that are in the $(k+2)$th level and in $N_B$. 
The working principle of this scheme with respect to the qubit $B$ is based on the fact that the tree cluster state is stabilized \cite{RB01} by operator $\hat{Z}_A \hat{X}_B \otimes_{i=1.\ldots b_{k+1}} \hat{Z}_{C^B_i} $ and the measurement outcome of the observable $\hat{Z}_A$ can thus be guessed by the parity of the observable $\hat{X}_B \otimes_{i=1.\ldots b_{k+1}} \hat{Z}_{C^B_i}$. 
Since the protocol with respect to another qubit $B' \in N_A$ in the same $(k+1)$th level works similarly and independently of that on the qubit $B$, 
the indirect $Z$-basis measurement on qubit $A$ succeeds 
when, under the parallel implementation of the schemes with respect to all the $(k+1)$th-level qubits in $N_A$, at least, one of them succeeds. 
In addition, the independence of the schemes allows us to use a majority vote  \cite{VBR06} to increase the fidelity of the measurement outcome of indirect $Z$-basis measurement on qubit $A$.
These properties are essential for greatly increasing the success probability of the loss-tolerant measurement as well as the robustness in the case of the measurement of observable $\hat{Z}$ or $\hat{X}$ against depolarization for the physical qubits (as seen in Secs.~\ref{se:error-rud} and \ref{se:num-rud}).

\subsection{Error analysis}\label{se:error-rud}

In our repeater scheme, we use the protocol of Varnava {\em et al.} to perform $Z$-basis or $X$-basis measurement under loss as well as general errors.
Thus, we assume the existence of individual depolarization for qubits in the tree cluster state with branching parameters $\{b_i\}_{i=0,1,\ldots,l}$,
and we investigate the effects of the errors for the loss-tolerant $Z$-basis and $X$-basis measurements under the use of a majority vote \cite{VBR06}.

The depolarizing channel ${\cal E}_A$ for qubit $A$ in state $\hat{\rho}$ is defined by
\begin{equation}
{\cal E}_A (\hat{\rho}) = (1- e_{\rm d}) \hat{\rho} + \frac{e_{\rm d}}{3} (\hat{X}_A \hat{\rho}\hat{X}_A +\hat{Y}_A \hat{\rho}\hat{Y}_A+\hat{Z}_A \hat{\rho}\hat{Z}_A),
\label{eq:dep}
\end{equation}
where $e_{\rm d}$ is the error probability of this channel. If we consider a Pauli measurement on the qubit $A$ in state ${\cal E}_A (\hat{\rho})$ (this is actually the case for our repeater scheme), the error probability $e_{\rm m}$ of this measurement is 
\begin{equation}
e_{\rm m}=\frac{2}{3} e_{\rm d}. \label{eq:em}
\end{equation}
Then, our error model considered in what follows can be specified as the one where all the qubits in the tree cluster state are subject to a depolarizing channel, independently,
and the goal here is to evaluate the effects of these errors for the loss-tolerant $Z$-basis and $X$-basis measurements.

We first consider the expectation value of the error probability $e_{{\rm I}_k}$ of the indirect $Z$-basis measurement on a qubit $A$ in the $k$th level in the tree.
As noted in the previous section, 
this measurement outcome of observable $\hat{Z}_A$ is guessed by the parity of the observable $\hat{X}_B \otimes_{i=1.\ldots b_{k+1}} \hat{Z}_{C^B_i}$, where $B$ is a $(k+1)$th-level qubit in $N_A $ and $\{C^B_i\}_{i=1.\ldots b_{k+1}}$ are $(k+2)$th-level qubits in $N_B$.
The success probability $S_k$ of the protocol to find out the parity of the observable $\hat{X}_B \otimes_{i=1.\ldots b_{k+1}} \hat{Z}_{C^B_i}$ is 
\begin{equation}
S_{k} = (1-\epsilon_0) (1-\epsilon_0 + \epsilon_0 R_{k+2})^{b_{k+1}}. \label{eq:S_k}
\end{equation}
This event requires the joint success of direct or indirect $Z$-basis measurements on all the qubits $\{C^B_i\}_{i=1.\ldots b_{k+1}}$. 
Suppose that, among the qubits $\{C^B_i\}_{i=1.\ldots b_{k+1}}$, 
$l_k $ qubits output only the outcomes of direct $Z$ measurements with error probability $e_{\rm m}$, while the other $(b_{k+1}-l_k)$ qubits present the measurement outcomes of indirect $Z$-basis measurements with average error probability $\bar{e}_{{\rm I}_{k+2}}$.
{\it Here we assume that the measurement outcome of indirect $Z$-basis measurement on a qubit is preferentially accepted if both of the direct and indirect measurement on it succeed}.
Then, in addition to the error probability of the direct $X$-basis measurement on qubit $B$, the average error probability of guessing the parity of the $\hat{X}_B \otimes_{i=1.\ldots b_{k+1}} \hat{Z}_{C^B_i}$, i.e., the average error probability $\bar{e}_{{\rm I}_k|B}$ of indirect $Z$-basis measurement on qubit $A$ from the measurements on the qubit $B$ and on the $(k+2)$th-level qubits around $B$, is estimated as
\begin{multline}
\bar{e}_{{\rm I}_{k}|B} = \sum_{l_k=0}^{b_{k+1}} 
\left(
\begin{array}{c}
b_{k+1} \\
l_k
\end{array}
\right)
\left( 1-\frac{R_{k+2} }{1-\epsilon_0 + \epsilon_0 R_{k+2} } \right)^{l_k}   \\
\times \left( \frac{R_{k+2} }{1-\epsilon_0 + \epsilon_0 R_{k+2} } \right)^{b_{k+1}-l_k} \\
\times \frac{1-(1-2 e_{\rm m})^{1+l_k } (1-2\bar{e}_{{\rm I}_{k+2}} )^{b_{k+1}-l_k} }{2}. \label{eq:err-i-b}
\end{multline}
This is the average error probability of the indirect $Z$-basis measurement scheme with respect to a qubit $B\in N_A$ in the $(k+1)$th level.
Note that we can also obtain the outcome of observable $Z_A$ by running such an indirect measurement scheme with respect to another qubit $B'\in N_A$ in the $(k+1)$th level.
Therefore, by performing all the schemes on all the $(k+1)$th-level qubits in $N_A$, 
we can take a majority vote over all the guessing outcomes of observable $Z_A$ that are obtained by all the successful ones. 
Suppose that there are $m_{k}(\ge 1)$ successful ones, which occurs with probability
\begin{equation}
T_k(m_k):=
\left(
\begin{array}{c}
b_k \\
m_k
\end{array}
\right) S_k^{m_k} (1-S_k)^{b_k-m_k}.
\end{equation}
Then, 
the average guessing probability $\bar{e}_{{\rm I}_{k}|m_k}$ of the measurement outcome of observable $\hat{Z}_A$ from the majority vote is
\begin{multline}
\bar{e}_{{\rm I}_{k}|m_k} =\\ 
\begin{cases}
    \sum_{j=\lceil m_k/2 \rceil}^{m_k} 
\left(
\begin{array}{c}
m_k \\
j
\end{array}
\right) (\bar{e}_{{\rm I}_{k}|B})^{j} (1-\bar{e}_{{\rm I}_{k}|B} )^{m_k-j},  \\
\hspace{4.7cm} (m_k\mbox{ is odd}), \\
       \sum_{j=\lceil m_k/2 \rceil}^{m_k-1} 
\left(
\begin{array}{c}
m_k-1 \\
j
\end{array}
\right) (\bar{e}_{{\rm I}_{k}|B})^{j} (1-\bar{e}_{{\rm I}_{k}|B} )^{m_k-1-j} \\ 
\hspace{4.7cm} (m_k\mbox{ is even}) ,
  \end{cases}
\end{multline}
where $\lceil x \rceil$ is the smallest integer among integers that are greater than or equal to $x$.
Therefore, the average error probability  $\bar{e}_{{\rm I}_{k}}$ is
\begin{equation}
\bar{e}_{{\rm I}_{k}}= \frac{1}{R_k} \sum_{m_k=1}^{b_k} T_k(m_k) \bar{e}_{{\rm I}_{k|m_k}},\label{eq:ex-ave}
\end{equation}
where we used
\begin{equation}
R_{k}= \sum_{m_k=1}^{b_k} T_k(m_k).
\end{equation}
$\{ \bar{e}_{{\rm I}_{k}} \}_{k=0,1,\ldots,l}$ can be derived by solving Eqs.~(\ref{eq:err-i-b})-(\ref{eq:ex-ave}) recursively.

Let us move to the error analysis for the loss-tolerant $Z$-basis measurement.
This measurement succeeds when direct or indirect $Z$-basis measurements on all the 1st-level qubits succeed (c.f., Fig.~\ref{fig:a2.eps}). Thus, the success probability $P_Z$ of the loss-tolerant $Z$-basis measurement is
described by
\begin{equation}
P_Z = (1-\epsilon_0 + \epsilon_0 R_1)^{b_0}. \label{eq:PZ}
\end{equation}
On the other hand, 
since all the 1st-level qubits are linked to all the qubits that have been connected to the encoded qubit as in Fig.~\ref{fig:a2.eps}, the average error probability $\bar{e}_Z$ of the loss-tolerant $Z$-basis measurement is 
\begin{multline}
\bar{e}_Z  = \sum_{l=0}^{b_{0}} 
\left(
\begin{array}{c}
b_{0} \\
l
\end{array}
\right)
\left( 1-\frac{R_{1} }{1-\epsilon_0 + \epsilon_0 R_{1} } \right)^{l} \\
\times \left( \frac{R_{1} }{1-\epsilon_0 + \epsilon_0 R_{1} } \right)^{b_{0}-l}
\frac{1-(1-2 e_{\rm m})^{l} (1-2\bar{e}_{{\rm I}_{1}} )^{b_{0}-l} }{2}. 
 \label{eq:ez}
\end{multline}

Let us proceed to the error analysis for the loss-tolerant $X$-basis measurement.
This measurement succeeds if direct $X$-basis measurement on any 1st-level qubit $B_1$ and direct or indirect $Z$-basis measurements on all the 2nd-level qubits $\{C_i^{B_1}\}_{i=1,\ldots,b_1}=:C(B_1)$ in $N_{B_1}$ succeed,
i.e., if we know the parity of $\hat{X}_{B_1} \hat{Z}_{C(B_1) }$, where $\hat{Z}_N:=\otimes_{i \in N} \hat{Z}_i$. 
To see this, let $\{B_i\}_{i=1,\ldots,b_0}=:B$ be 1st-level qubits, 
let $\{A_i \}_{i}=:A$ be qubits that have been connected to the encoded qubit, i.e., $A = N_{B_i} \setminus C(B_i)$ (c.f., Fig.~\ref{fig:a2.eps}).
Through the measurement of  parity $k_1(=0,1)$ of observable $\hat{X}_{B_1} \hat{Z}_{C(B_1)} $, the stabilizers for the initial cluster state are renewed as 
\begin{align}
\left\{
\begin{array}{l}
\hat{X}_{A_i} \hat{Z}_{B} \hat{Z}_{N_{A_i} \setminus B}, \\
\hat{Z}_{A} \hat{X}_{B_i} \hat{Z}_{C(B_i)} ,
\end{array}
\right.
\to
\left\{
\begin{array}{l}
\hat{X}_{A_1} \hat{X}_{A_i} \hat{Z}_{N_{A_1}\setminus B } \hat{Z}_{N_{A_i} \setminus B} \;\; (i\neq 1) ,\\
(-1)^{k_1} \hat{Z}_{A} ,   \\
(-1)^{k_1}\hat{X}_{B_1} \hat{Z}_{C(B_1)}, \\
(-1)^{k_1} \hat{X}_{B_i} \hat{Z}_{C(B_i)} \;\; (i \neq 1). 
\end{array}
\right.
\end{align}
The stabilizers in the first and second rows of the right-hand side (RHS) of this equation correspond to the desired backaction that is the same as that of the direct $X$-basis measurement on the encoded qubit.
The stabilizers in the last row of the RHS indicate that
$\{B_i\}_{i=2,\ldots,b_0}$ are decoupled from qubits in $A$ and the parity of observable $\hat{X}_{B_i} \hat{Z}_{C(B_i)}$ $(i \neq 1)$ is the same as $k_1$, 
These facts suggest that the loss-tolerant $X$-basis measurement succeeds if we succeed in finding out one of the parities of $\{ \hat{X}_{B_i} \hat{Z}_{C(B_i)}\}_{i=1,\ldots,b_0}$,
and that we can use a majority vote like the indirect $Z$-basis measurement on a qubit in a level.
Therefore, the success probability $P_X$ of the loss-tolerant $X$-basis measurement is
\begin{equation}
P_X=R_0,\label{eq:PX}
\end{equation}
and the average error probability is
\begin{equation}
\bar{e}_X= \bar{e}_{{\rm I}_0}.\label{eq:ex}
\end{equation}

Note that the success probability of the loss-tolerant measurement for general observable $\hat{A} (\alpha)$ is less than those for observables $\hat{Z}$ and $\hat{X}$,
i.e., $P_{\rm L} \le P_Z$ and $P_{\rm L} \le P_X$.
In addition, the average error probability $\bar{e}_{\rm L}$ of the loss-tolerant measurement for general observable $\hat{A}(\alpha)$ is in the order of $e_{\rm m}$ owing to the direct contribution of the error of direct measurement of observable $\hat{A}(\alpha)$ on a 1st-level qubit \cite{VBR06}.
This is in contrast to the error probabilities $\bar{e}_Z$ and $\bar{e}_X$
that have a mechanism---a majority vote---to greatly suppress the error induced by individual depolarization. 
{\it These contrasts in error probability and success probability between the protocol for observable $\hat{Z}$ or $\hat{X}$ and that for general $\hat{A} (\alpha)$ represent a notable difference between quantum repeaters and quantum computing,
in the sense that quantum repeaters are possible without such a general single-qubit measurement like our protocol but quantum computation is not} \cite{GK}.

\subsection{Numerical examples}\label{se:num-rud}

To ensure the special robustness against general errors that appears in the cases of $Z$-basis and $X$-basis measurements,
here we present numerical examples of $P_Z$, $P_X$, $P_{\rm L}$, $\bar{e}_Z$, and $\bar{e}_X$.
(I) For $\epsilon_0 \simeq 0.20$ and $e_{\rm m}=2e_{\rm d}/3=2.8 \times 10^{-5}$, 
the encoded qubit with branching parameters $\{b_0,b_1,b_2\}=\{16,14,1\}$ ($\{b_0,b_1,b_2\}=\{11,11,1\}$), which is composed of $Q_{\rm L}=464$ ($Q_{\rm L}=253$) qubits, gives $1-P_Z=1.8\times 10^{-6}$ ($1-P_Z=2.5\times 10^{-5}$),
$1-P_X=5.9\times 10^{-5}$ ($1-P_X=3.2\times 10^{-4}$),
$1-P_{\rm L}=0.43$ ($1-P_{\rm L}=0.35$),
$\bar{e}_Z=1.6 \times 10^{-7}$ ($\bar{e}_Z=1.4 \times 10^{-6}$),
and $\bar{e}_X=2.5 \times 10^{-6}$ ($\bar{e}_X=8.3 \times 10^{-6}$).
(II) For $\epsilon_0  \simeq 0.27$ and $e_{\rm m}=2e_{\rm d}/3=5.6 \times 10^{-5}$, 
the encoded qubit with branching parameters $\{b_0,b_1,b_2\}=\{17,28,2\}$ ($\{b_0,b_1,b_2\}=\{12,23,2\}$), which is composed of $Q_{\rm L}=1445$ ($Q_{\rm L}=840$) qubits, gives $1-P_Z=4.9\times 10^{-6}$ ($1-P_Z=3.7\times 10^{-5}$),
$1-P_X=1.1\times 10^{-4}$ ($1-P_X=5.6\times 10^{-4}$),
$1-P_{\rm L}=0.43$ ($1-P_{\rm L}=0.37$),
$\bar{e}_Z=5.2 \times 10^{-7}$ ($\bar{e}_Z=2.8 \times 10^{-6}$),
and $\bar{e}_X=1.5 \times 10^{-5}$ ($\bar{e}_X=4.7 \times 10^{-5}$).
These data sets (I) and (II) are obtained by numerical calculation based on Eqs.~(\ref{eq:var-1})-(\ref{eq:rk}), (\ref{eq:PZ}), (\ref{eq:ez}), 
(\ref{eq:PX}), and (\ref{eq:ex}),
and they will be used in Sec.~\ref{se:num} to estimate the performance of our repeater protocol for $L_0=4$ km and $L_0=8$ km.
For those choices of the parameters, the success probabilities $P_Z$ and $P_X$ for $Z$-basis and $X$-basis measurements are {\it much higher} than the success probability $P_{\rm L}$ for general observable $\hat{A}(\alpha)$.
More specifically, the branching parameters chosen here are too small to obtain $P_{\rm L}> 1- \epsilon_0$, but they are large enough to achieve our quantum repeater protocol that uses only the loss-tolerant $Z$-basis or $X$-basis measurements.
In addition, the {\it smaller} values of $\bar{e}_Z$ and $\bar{e}_X$ than $e_{\rm m}$ imply that the loss-tolerant $Z$-basis or $X$-basis measurement is robust even against individual errors on physical qubits, which is in contrast to the general observable $\hat{A}(\alpha)$.
We also note that the greater the branching ratios the trees have, the lower the failure probability and error probability of the loss-tolerant $Z$-basis or $X$-basis measurement \cite{VBR06} become.

\section{Preparation of an encoded complete-like cluster state $\ket{\bar{G}_{\rm c}^m}$}\label{se:preparation}

\begin{figure}[b]  
\includegraphics[keepaspectratio=true,height=25mm]{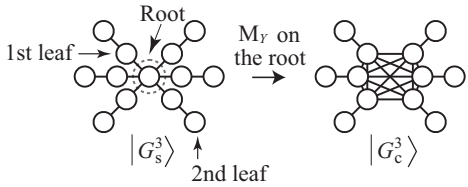}
\caption{Star-like cluster state $\ket{G_{\rm s}^m}$ (for the case of $m=3$). $Y$-basis measurement M$_Y$ on the root qubit transforms the state into a complete-like cluster state $\ket{G_{\rm c}^m}$, up to local unitary operations \cite{HDERVB}.}
  \label{fig:a3.eps}
\end{figure}

In this section, we estimate the resources and the time required to prepare an encoded version $\ket{\bar{G}_{\rm c}^m}$ of a complete-like cluster state in Fig.~\ref{fig:3.eps} with linear optical elements, single-photon sources, photon detectors, and a high-speed active feedforward technique.
The preparation time estimated here will be translated into the corresponding loss for photons in our repeater protocol discussed in Sec.~\ref{se:repeater}. 
Here we begin by considering the preparation of an encoded {\it star-like} cluster state $\ket{\bar{G}_{\rm s}^m}$ that can be transformed into the state $\ket{\bar{G}_{\rm c}^m}$.

The star-like cluster state denoted by $\ket{G_{\rm s}^m}$ is described by the graph of Fig.~\ref{fig:a3.eps}. 
This state centers a single root qubit that has $2 m$ arms ($m \ge 1$) composed of 1st-leaf and 2nd-leaf qubits.
The state has been used as the basic unit to produce 2-dimensional or 3-dimensional cluster states \cite{CCWD06,FT10}.
The star-like cluster state with 1st-leaf qubits being encoded is denoted by $\ket{\bar{G}_{\rm s}^m}$, and the encoding allows us to execute the loss-tolerant measurements on the 1st-leaf qubits.
In particular, the state $\ket{\bar{G}_{\rm s}^m}$ is obtained by replacing the 1st-leaf qubits in the state $\ket{G_{\rm s}^m}$ with the root qubits of the $Q_{\rm L}$-qubit tree cluster states with branching parameter $\{b_i\}_{i=0,1,\ldots,l}$, where
the root and 0th-level qubits of the $Q_{\rm L}$-qubit tree cluster states are to receive $X$-basis measurements in order to complete the loss-tolerant encoding.
We note that the whole state $\ket{\bar{G}_{\rm s}^m}$ is (not equivalent to but) similar to a tree cluster state with branching parameters $\{2m,2,b_0,\ldots,b_l \}$. Thus, it can be prepared efficiently via the protocol \cite{VBR07,VBR08} of Varnava {\em et al.} 

The protocol of Varnava {\em et al.} is \cite{VBR07,VBR08} based on the single-photon sources with efficiency $\eta_{\rm S}$, linear optical elements, and single-photon detectors with quantum efficiency $\eta_{\rm D}$, and it proceeds as follows:
(a) We first prepare a three-qubit GHZ state with an effective individual loss probability $1-\eta_{\rm S}/(2-\eta_{\rm D} \eta_{\rm S})$ via a protocol 
that uses 6 photons for each trial and
succeeds with probability $\eta_{\rm D}^3 \eta_{\rm S}^3(2-\eta_{\rm D} \eta_{\rm S})^3/32$.
(b) Then, we produce a``2-tree'' by applying a ``type-II fusion'' to a pair of the three-qubit GHZ states.
(c) We then create the target tree cluster state $\ket{\bar{G}_{\rm s}^m}$ (whose root qubit is redundantly encoded) from the 2-trees by using the type-II fusions.
It follows that the expected number $\bar{Q}_{\rm s}$ of total single photons required to produce the state $\ket{\bar{G}_{\rm s}^m}$ (whose root qubit is actually a redundantly-encoded qubit composed of two bare qubits \cite{VBR07,VBR08}) is  \cite{VBR07,VBR08} bounded by
\begin{align}
\bar{Q}_{\rm s} \le \frac{2\times 6 \times 32}{\eta_{\rm S}^3 \eta_{\rm D}^3 (2-\eta_{\rm S} \eta_{\rm D})^3}  \frac{1}{P_{\rm II}^{2l+4}} {\rm poly}(2m) \prod_{i=0}^l {\rm poly} (b_i), \label{eq:qsup}
\end{align}
where $P_{\rm II}$ is the success probability of the type-II fusion.
On the other hand, if we assume that  $\tau_{\rm a}$ represents the time to execute a two-qubit or single-qubit measurement and the associated classical feedforward,
the time $\tau_{\rm s}$ needed to prepare the state $\ket{\bar{G}_{\rm s}^m}$ is \cite{VBR07,VBR08} described by
\begin{align}
\tau_{\rm s}\simeq & \left(\log_2 2m +  \sum_{i=0}^l \log_2 b_i +l + 2 + 2 \right) \tau_{\rm a} \nonumber \\
=& \left(\log_2 2m +  \sum_{i=0}^l \log_2 b_i +l + 4 \right) \tau_{\rm a}, \label{eq:tau_s}
\end{align}
where the last term $2 \tau_{\rm a}$ in the first equation comes from steps (a) and (b) and the other terms in the same equation result from step (c).
The expression of $P_{\rm II}$ in Eq.~(\ref{eq:qsup}) actually depends on the timing of the application of the corresponding type-II fusion gate,
but it is at least bounded as
\begin{equation}
\frac{\eta_{\rm S}^2 \eta_{\rm D}^2 P_{\tau_{\rm a} }^{2 \tau_{\rm s} / \tau_{\rm a}}} {2(2-\eta_{\rm S} \eta_{\rm D})^2} \le P_{\rm II} \le  \frac{\eta_{\rm S}^2 \eta_{\rm D}^2 } {2(2-\eta_{\rm S} \eta_{\rm D})^2},
\end{equation}
where $P_{\tau_{\rm a}}$ is the survival probability of a photon for the time $\tau_{\rm a}$.
Note that 
the bound of Eq.~(\ref{eq:qsup}) scales polynomially with the total photon number of the state $\ket{\bar{G}_{\rm s}^m}$, i.e., $2m(Q_{\rm L}+3)+2$.
Through this preparation, the effective loss probability of the individual photons in the state $\ket{\bar{G}_{\rm s}^m}$ is $1-P_{\tau_{\rm a} }^{ \tau_{\rm s} / \tau_{\rm a}} \eta_{\rm S}/(2 -\eta_{\rm D}  \eta_{\rm S}) $.

As shown in Fig.~\ref{fig:a3.eps}, the encoded complete-like cluster state $\ket{\bar{G}_{\rm c}^m}$ can be prepared by performing the (effective) $Y$-basis measurement on the (redundantly-encoded \cite{VBR07,VBR08}) root qubit of the state $\ket{\bar{G}_{\rm s}^m}$ as well as proper local unitary operations [note that the $Y$-basis measurement corresponds to a Bell measurement (on a two-dimensional subspace)]. 
We also need to perform $X$-basis measurements on the 1st-leaf qubits (corresponding to the root qubits of the $Q_{\rm L}$-qubit tree cluster states) in the state $\ket{\bar{G}_{\rm s}^m}$ as well as on the associated 0-level qubits in the tree, in order to complete the encoding of Fig.~\ref{fig:a2.eps}.
These measurements multiply an additional factor 
$
[ (2-\eta_{\rm S} \eta_{\rm D})/(\eta_{\rm S} \eta_{\rm D} P_{\tau_{\rm a} }^{\tau_{\rm s} / \tau_{\rm a}} )]^2 \times
[ (2-\eta_{\rm S} \eta_{\rm D})/(\eta_{\rm S} \eta_{\rm D} P_{\tau_{\rm a} }^{\tau_{\rm s} / \tau_{\rm a}} )]^{4m}
$
to $\bar{Q}_{\rm s}$,
where the first term and the second term come from the $Y$-basis measurement and from the $X$-basis measurements, respectively.
Therefore, the expected number $\bar{Q}_{\rm c}$ of total single photons needed to produce the state $\ket{\bar{G}_{\rm c}^m}$ and the time $\tau_{\rm c}$ are 
\begin{align}
\bar{Q}_{\rm c} =&  \left[\frac{2-\eta_{\rm S} \eta_{\rm D}}{\eta_{\rm S} \eta_{\rm D} P_{\tau_{\rm a} }^{\tau_{\rm s} / \tau_{\rm a}} } \right]^{4m+2} \bar{Q}_{\rm s} ,\\
\tau_{\rm c} =& \tau_{\rm s}+ \tau_{\rm a} .\label{eq:tau_c}
\end{align}
Note that, since the preparation of the state $\ket{\bar{G}_{\rm c}^m}$ is executed locally, the total photon number required to produce the state $\ket{\bar{G}_{\rm c}^m}$ is independent of the distance between Alice and Bob in our repeater protocol of Fig.~\ref{fig:3.eps}.

\section{Scaling of our quantum repeater protocol}\label{se:repeater}

Let us discuss the performance of our all photonic quantum repeater protocol defined in Fig.~\ref{fig:3.eps}.
Let $\epsilon_0$ be an effective probability with which the photon loss occurs in the local preparation of the state $\ket{\bar{G}_{\rm c}^m}$ 
or in the transmission of photons between adjacent source and receiver nodes, 
which is selected to satisfy $\epsilon_0 <0.5$.
We also assume that every photon is subject to individual depolarization in the transmission from a sender node to the adjacent receiver node.
Let $e_{\rm d} $ be the error probability caused by this depolarization [c.f., Eq.~(\ref{eq:dep}) and the relation with $e_{\rm m}$].

In Sec.~\ref{se:tot},
we present the average total photon number $\bar{Q}$ consumed in our protocol to produce an entangled pair between Alice and Bob, the scaling of $\bar{Q}$, 
and the average rate $\bar{R}$ of our protocol.
In addition, in the section, we show a flexibility of our protocol through considering the memory time that is needed in the case of the extreme application, i.e., quantum teleportation \cite{B93}.
In Sec.~\ref{se:error}, we analyze the average errors for the obtained entangled pair and give the scaling.
Section \ref{se:num} presents numerical examples of these quantities as well as a comparison with the speediest protocol given by Munro {\it et al.} \cite{M12}.

\subsection{Total photon number and entanglement generation rate of our repeater protocol} \label{se:tot}

Here we first present the performance of our repeater protocol that is characterized by the success probability, the average rate $\bar{R}$ for producing an entangled pair between Alice and Bob, and the average total photon number $\bar{Q}$ consumed in the protocol to produce the pair.
Next, we show a flexibility of our protocol via considering the requirements for our repeater protocol that emerge when it is used to accomplish quantum teleportation \cite{B93}.
Finally, we derive the scaling of $\bar{Q}$.

Let us begin by considering the performance of our repeater protocol.
Suppose that our protocol defined in Fig.~\ref{fig:3.eps} succeeds. In this case, single photons in a linear cluster state connecting Alice's qubit $A$ and Bob's qubit $B$ are presented, which then receive $X$-basis measurements at all the receiver nodes. Thus, according to the rule of Fig.~\ref{fig:2.eps}(c), Alice and Bob's qubits $AB$ are entangled.
Since the success events occur when all the receiver nodes have at least one successful Bell measurement and all the loss-tolerant measurements on the 1st-leaf qubits succeed, the total success probability $P$ is described by
\begin{equation}
P= P_Z^{2 (m-1)n} P_X^{2 n} [1- (1-P_{\rm B})^m]^{n+1}, \label{eq:P}
\end{equation}
where $P_Z$ is the success probability of the loss-tolerant $Z$-basis measurement [c.f., Eq.~(\ref{eq:PZ})], 
$P_X$ is the success probability of the loss-tolerant $X$-basis measurement [c.f., Eq.~(\ref{eq:PX})], and
$P_{\rm B}$ is the success probability of the Bell measurement
represented by 
\begin{equation}
P_{\rm B} =\frac{(1-\epsilon_0)^2}{2} .
\end{equation}
By using the success probability $P$, the average rate $\bar{R}$ to produce an entangled pair between Alice and Bob is simply described as
\begin{equation}
\bar{R}=Pf, \label{eq:tau_QKD}
\end{equation}
where $f$ is the repetition rate of the slowest device among single-photon sources, photon detectors, and active-feedforward techniques.
Note that we can increase this rate as high as we want just by running single-photon sources as fast as required {\it rather than by increasing the number of the devices and using them in parallel}.
This is in striking contrast to the standard quantum repeater protocols \cite{B98,DLCZ,SSRG09,KWD03,C06,L06,La06,K08,ATKI10,ZDB12,L12,S07,A09,A10,M08,L08,R09} as in Fig.~\ref{fig:1.eps},
where we cannot adopt such a machine-gun-like style because the protocols, at least, need to receive the heralding signals for entanglement swapping from {\it distant} nodes.
On the other hand, the average of the total number $Q$ of single photons consumed in our repeater protocol to produce the entangled pair is
\begin{equation}
\bar{Q}= \frac{2mn (Q_{\rm L}+1) +2m}{P}. \label{eq:Q}
\end{equation}
In general, 
{\it it would be better to choose all the parameters so as to minimize $\bar{Q}$}.

Note that the maximum time $t_{\rm max}$ while photons in our protocol are expected to survive is 
\begin{equation}
t_{\rm max}=\tau_{\rm c} + \frac{L_0}{2c}+2 \tau_{\rm a}, \label{eq:tmax}
\end{equation}
where $\tau_{\rm c}$ is the preparation time of the state $\ket{\bar{G}_{\rm c}^m}$,
$L_0/(2c)$ is the transmission time of photons from a source node to the adjacent receiver node, and
$2\tau_{\rm a}$ in this expression comes from the Bell measurement in step (ii) of our protocol in Fig.~\ref{fig:3.eps} and from the loss-tolerant measurements on the 1st-leaf qubits in the step (iii).
Combined with the results in Sec.~\ref{se:preparation} and with a fact that every photon is finally fed into photon detectors with quantum efficiency $\eta_{\rm D}$,
the effective total loss probability $\epsilon_0$ is approximately described by
\begin{equation}
\epsilon_0 \simeq 1-e^{-L_0/(2 l_{\rm att})} P_{\tau_{\rm a} }^{ (\tau_{\rm c} +2 \tau_{\rm a})/ \tau_{\rm a}} \frac{\eta_{\rm D} \eta_{\rm S}}{2 -\eta_{\rm D}  \eta_{\rm S}}, \label{eq:epsilon_0}
\end{equation}
where $l_{\rm att}$ is the attenuation length of the optical channel between a source node and the adjacent receiver node.

Next, let us see a flexibility of our protocol through
considering the requirements for our protocol in the case of the application to quantum teleportation \cite{B93}. 
If we use our protocol for quantum key distribution (QKD), the protocol does not require any quantum memory as noted in the main body of our paper.
However, 
when we use our protocol to achieve the most general quantum communication based on quantum teleportation \cite{B93},
the communicators, Alice and Bob, should have quantum memories,
because quantum teleportation protocol itself requires \cite{B93} the sender (Alice) to send the outcome of her Bell measurement to the receiver (Bob).
But, depending on the coherence time of their quantum memories, 
they have two choices that make a difference in the required memory time $t_{\rm mem}$ for achieving the quantum teleportation:
(a) To keep the memory time $t_{\rm mem}$ to the minimum, i.e., the classical communication time $L/c$ required in the quantum teleportation, 
they can set $P \simeq 1$ through the consumption of much more single photons (i.e., by increasing $\bar{Q}$), which leads to
\begin{equation}
t_{\rm mem}=\frac{L_0}{2c}+2 \tau_{\rm a}+\frac{L}{c} , \label{eq:tau_tele-1}
\end{equation}
where the first two terms are the time needed in our repeater to prepare an entangled pair.
As another strategy, (b) they can start quantum teleportation,
upon receiving the signals for heralding the success of the repeater protocol from receiver nodes, making the required memory time $t_{\rm mem}$ be
\begin{equation}
t_{\rm mem}=\frac{L_0}{2c}+2 \tau_{\rm a}+\frac{L-L_0/2}{c} + \frac{L}{c}=2 \tau_{\rm a}+\frac{2 L }{c} , \label{eq:tau_tele}
\end{equation}
where the additional term $(L-L_0/2)/c$ is the classical communication time between Alice and the receiver node adjacent to Bob for confirming the success of the repeater protocol.
The strategy (a) is effective when the coherence time of the quantum memories is short,
while the strategy (b) has a merit that the total number $\bar{Q}$ of single photons consumed in our repeater protocol can be set to the minimum.
However, in most of the cases, it would be better to adopt the strategy (b) from the following sense:
If Alice and Bob try to accomplish quantum teleporation, 
they must have quantum memories with coherence time longer than classical communication time between them (that is required in the quantum teleportation protocol) at least, $L/c$, as in Eq.~(\ref{eq:tau_tele-1}).
Thus, in such an era when Alice and Bob want to achieve quantum teleportation,
the additional coherence time  $(L-L_0/2)/c$ in Eq.~(\ref{eq:tau_tele}) must not be the problem for them,
implying that it would be better to adopt the strategy (b).

However, we stress that, if we use a conventional quantum repeater protocol \cite{B98,DLCZ,SSRG09,KWD03,C06,L06,La06,K08,ATKI10,ZDB12,L12,S07,A09,A10,M08,L08,L09,R09} based on quantum memories as described in Fig.~\ref{fig:1.eps},
the required memory times $t_{\rm mem}$ include, at least, further additional factors that come from the classical communication time to trigger the next round of entanglement swapping as shown in Fig.~\ref{fig:1.eps}, 
irrespectively of the application (QKD or quantum teleportation).
In addition, in the standard quantum repeater protocols for $L=1000$~km,
$t_{\rm mem} $ has been estimated \cite{DLCZ,SSRG09,ATKI10,C06}, at least, 1~s.
This is in a striking contrast to our protocol that suppresses the memory time $t_{\rm mem}$, which required only in the case of quantum teleportation, to about $10$~ms even in the strategy (b) for  $L=1000$~km.
These facts suggest an advantage to use our protocol instead of the conventional quantum repeaters even if we enter an era when we have a matter quantum memory,
in the sense that the smaller $t_{\rm mem}$ of our protocol should lead to smaller errors (depolarization or dephasing) of the quantum memories.
Moreover, $t_{\rm mem}$ of our protocol with the strategy (a) becomes the same as the speediest repeater protocol \cite{M12} based on matter qubits at repeater nodes.
Therefore, our protocol is flexible, and it has a distinguished advantage over the standard quantum repeater protocols with respect to the required memory time $t_{\rm mem}$.

We further note that the flexibility seen in the strategies (a) and (b) extends the range of the applications of our all photonic quantum repeaters.
For example, even if we want to use our protocol as an almost deterministic entanglement supplier for accomplishing another quantum information processing protocols, such as nonlocal measurements \cite{V03,C10} and cheating strategies \cite{LL11,KMS11} in position-based quantum cryptography \cite{B11}, {\it in a memory-less fashion},  
we can achieve this, at least, by adopting strategy (a). 
Thus, the flexibility makes any kind of quantum communication scheme possible with our all photonic quantum repeater system alone.

\begin{figure*}[tb]  
\includegraphics[keepaspectratio=true,height=22mm]{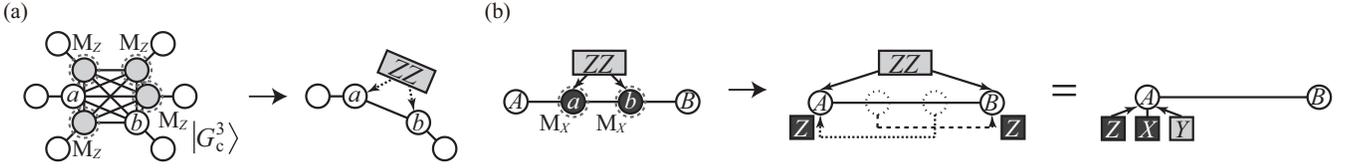}
\caption{(a) Error propagation of $Z$-basis measurements on $2(m-1)$ 1st-leaf qubits in state $\ket{G_{\rm c}^m}$ ($m=3$). Depending on the total parity $k$ of the $Z$-basis measurements, unitary operation $(\hat{Z}_{ab})^k$ ($\hat{Z}_{ab}:=\hat{Z}_a \hat{Z}_b$) is performed. The error of the $Z$-basis measurements thus leads to a phase-flip channel $\Lambda^{Z_{ab}}$.
(b)  Error propagation of two adjacent loss-tolerant $X$-basis measurements on qubits $ab$ that connect two-end qubits $AB$ linearly.
The measurement outcome $k_a$ ($k_b$) on qubit $a$ ($b$)  is correlated with the phase flip $(\hat{Z}_B)^{k_a}$ ($(\hat{Z}_A)^{k_b}$),
implying that the measurement error leads to the phase-flip channel on qubit $B$ ($A$).
Since the stabilizers for a bipartite cluster state are $\hat{X}_A \hat{Z}_B$ and $\hat{Z}_A \hat{X}_B$, $\hat{Z}_B$ ($\hat{Z}_{AB}$) has the same action with $\hat{X}_A$ ($\hat{Y}_A$) for the state.
}
  \label{fig:2}
\end{figure*}

Finally, let us derive the scaling of $\bar{Q}$ for the distance $L$ between Alice and Bob.
Here, instead of $\bar{Q}$, for simplicity, we consider the scaling of an upper bound from $P_X \ge P_{\rm L} $ and $P_Z \ge P_{\rm L}$,
\begin{equation}
\bar{Q}_{\rm upp}:= \frac{2mn (Q_{\rm L}+1) +2m}{P_{\rm L}^{2mn} [1- (1-P_{\rm B})^m]^{n+1}} (\ge \bar{Q}),
\end{equation}
for a specific choice of the parameters, 
\begin{align}
&m \simeq  \log_{1-P_{\rm B}} \left[ \frac{2 n \ln P_{\rm L}}{2 n \ln P_{\rm L} + (n+1) \ln (1-P_{\rm B})} \right], \\
&P_{\rm L} = (1-P_{\rm B})^{\frac{x}{n}},
\end{align}
where $x$ is a positive constant.
Note that we can choose any $x>0$ as long as $\epsilon_0<0.5$. 
Then, combined with Eq.~(\ref{eq:QI}), the upper bound $\bar{Q}_{\rm upp}$ becomes
\begin{multline}
\bar{Q}_{\rm upp} \simeq  2
 \left[n \left( \left[  \ln \frac{1}{1-(1-P_{\rm B})^{\frac{x}{n}}} \right]^{4.5} +1\right) +1 \right] \\
\times \left( 1+\frac{n+1}{2x}\right)^{2x} \left( 1+\frac{2x}{n+1} \right)^{n+1} \\
\times  \log_{1-P_{\rm B}} \left(  \frac{2x}{2x+(n+1)} \right).
\end{multline}
Since $n=L/L_0-1$, it is concluded that the average number $\bar{Q}$ of the total single photons required to produce an entangled pair scales only polynomially with distance $L$.

\subsection{Error analysis for our repeater protocol}\label{se:error}

Here we consider the error probabilities for the obtained entangled pair $AB$ in the success case of our repeater protocol.
We further derive the scaling of these error probabilities.

In the success case of our repeater protocol, $2(m-1)$ 1st-leaf qubits in the encoded complete-like cluster state $\ket{\bar{G}_{\rm c}^m}$ receive loss-tolerant $Z$-basis measurements with average error probability $\bar{e}_Z$.
We start by considering the effect of the error that comes from this measurement process.
Suppose that we perform $Z$-basis measurement on a qubit $A$ in a cluster state.
Then, note that the cluster state is stabilized by operators $\hat{X}_A \hat{Z}_{N_{A}}$ and $\hat{X}_i \hat{Z}_A \hat{Z}_{N_i\backslash A} $ for $i \in N_A$ \cite{RB01},
where $\hat{Z}_{N}:=\otimes_{i\in N } \hat{Z}_i $ for a set $N$ of qubits.
Through the $Z$-basis measurement on qubit $A$, depending on its outcome $k(=0,1)$, these stabilizers are renewed as $(-1)^{k} \hat{Z}_A$ and  $(-1)^k \hat{X}_i \hat{Z}_{N_i\backslash A} $ for $i \in N_A$, which stabilize the cluster state after the measurement. This fact implies that the measurement outcome $k$ is correlated with the application of unitary $(\hat{Z}_{N_A})^k$.
Thus, non-zero error probability $e_{Z}$ of the $Z$-basis measurement on qubit $A$ 
leads to a channel $\Lambda^{Z_{N_A}}_{1-2e_Z}$, where 
\begin{align}
\Lambda^{W_C}_{1-2e} (\hat{\rho}) :=& (1-e) \hat{\rho} + e \hat{W}_C \hat{\rho} \hat{W}_C \nonumber \\
 =& \frac{1+(1-2e)}{2}\hat{\rho} + \frac{1-(1-2e)}{2}\hat{W}_C \hat{\rho} \hat{W}_C
\end{align}
for a Pauli operator $\hat{W}_C$ on system $C$.
As for the loss-tolerant $Z$-basis measurements on $2(m-1)$ 1st-leaf qubits in state $\ket{\bar{G}_{\rm c}^m}$,  
the effect of the measurement error $\bar{e}_Z$ is described by phase-flip channel $\Lambda^{Z_{ab}}_{(1-2\bar{e}_Z)^{2m-2}}$ on average,
where $a$ and $b$ are the remaining 1st-leaf two qubits.
This result is summarized as in Fig.~\ref{fig:2}(a).

Next, we consider error propagation caused by two adjacent loss-tolerant $X$-basis measurements with error probability $\bar{e}_X$ on the remaining 1st-leaf qubits $ab$ (c.f., Fig.~\ref{fig:2}(b)).
Here let us call the 2nd-leaf qubits $A$ and $B$ as defined in Fig.~\ref{fig:2}(b).
We start by noting that the initial state is stabilized by operators $\hat{X}_{A}\hat{Z}_a$, $\hat{Z}_{A}\hat{X}_{a}\hat{Z}_{b}$,
 $\hat{Z}_{a}\hat{X}_{b}\hat{Z}_{B}$, and $\hat{Z}_{b}\hat{X}_{B}$.
When we assume to obtain outcomes $k_a(=0,1)$ and $k_b(=0,1)$ for the respective ideal $X$-basis measurements on qubits $a$ and $b$, 
these stabilizers are converted to $(-1)^{k_a} \hat{X}_a$,
 $(-1)^{k_{b}} \hat{X}_{b}$, $(-1)^{k_{b}} \hat{X}_{A}\hat{Z}_{B}$, and $(-1)^{k_a}\hat{Z}_{A}\hat{X}_{B}$. This implies that the measurement outcome $k_a$ ($k_{b}$) is correlated with the application of unitary $(\hat{Z}_{B})^{k_{a}}$ ($(\hat{Z}_{A})^{k_{b}}$). Thus, the non-zero average error probability $\bar{e}_{X}$ of the loss-tolerant $X$-basis measurement on qubit $a$ ($b$)
leads to a phase-flip channel $\Lambda^{Z_{B}}_{1-2\bar{e}_X}$ ($\Lambda^{Z_{A}}_{1-2\bar{e}_X}$), and the inherent phase-flip channel $\Lambda^{Z_{ab}}_{(1-2\bar{e}_Z)^{2m-2}}$ on qubits $ab$ becomes phase-flip channel $\Lambda^{Z_{AB}}_{(1-2\bar{e}_Z)^{2m-2}}$ on qubits $AB$.
Since these channels are considered to be applied to a {\it bipartite} cluster state,
as noted in Fig.~\ref{fig:2}(b),
the effects of $\Lambda^{Z_{B}}_{1-2\bar{e}_X}$ and $\Lambda^{Z_{AB}}_{(1-2\bar{e}_Z)^{2m-2}}$ are equivalent to $\Lambda^{X_{A}}_{1-2\bar{e}_X}$ and $\Lambda^{Y_{A}}_{(1-2\bar{e}_Z)^{2m-2}}$, respectively.

\begin{figure*}[t]  
\includegraphics[keepaspectratio=true,height=39mm]{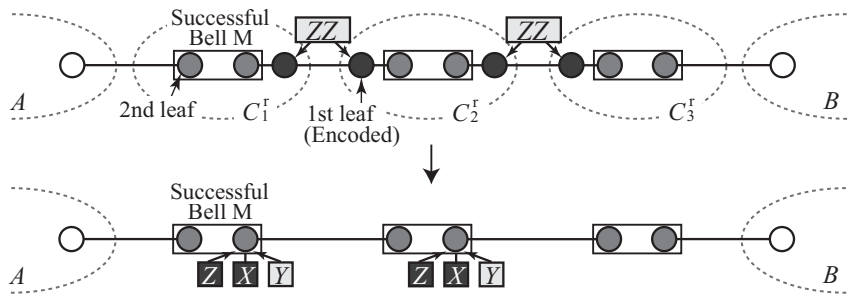}
\caption{Error propagation in the succees case of our repeater protocol defined in Fig.~\ref{fig:3.eps} ($n=2$). The Bell measurement here is defined in Fig.~\ref{fig:2.eps}(b). The 2nd-leaf qubits are subject to the depolarization represented by ${\cal E}$ (without exception), and we describe this effect by coloring the 2nd-leaf qubits in gray.  
}
  \label{fig:3}
\end{figure*}

Finally, we derive error probabilities of a Bell pair $AB$ that is obtained by the success of our repeater protocol defined in Fig.~\ref{fig:3.eps}. 
In the case of the success, the protocol can be regarded as a situation like the upper one of Fig.~\ref{fig:3} where the encoded complete-like cluster state $\ket{\bar{G}_{\rm c}^m}$ has already been transformed into a 4-qubit linear cluster state through the $Z$-basis measurements as in Fig.~\ref{fig:2}(a), and the 2nd-leaf qubits in the 4-qubit linear cluster state are connected via the successful Bell measurement defined in Fig.~\ref{fig:2.eps}(b).
This situation can further be transformed into the the lower one of Fig.~\ref{fig:3}, according to the rule of Fig.~\ref{fig:2}(b).
Therefore, considering the depolarization ${\cal E}$ for the 2nd-leaf qubits,
we conclude that
the finally obtained entangled pair $AB$ in a cluster state has errors specified by channel
\begin{align}
\bar{{\cal E}}^{\rm tot}_{A} := & (\Lambda^{Z_{A}}_{1-2\bar{e}_X})^n (\Lambda^{X_{A}}_{1-2\bar{e}_X})^n  (\Lambda^{Y_{A}}_{(1-2\bar{e}_Z)^{2m-2}})^{n} {\cal E}_A^{2(n+1)} \nonumber \\
=& \Lambda^{Z_{A}}_{(1-2\bar{e}_X)^n} \Lambda^{X_{A}}_{(1-2\bar{e}_X)^n} \Lambda^{Y_{A}}_{(1-2\bar{e}_Z)^{(2m-2)n}} {\cal E}_A^{2(n+1)},
\end{align}
on average. 
If we rewrite $\bar{{\cal E}}^{\rm tot}_{A}$ in a standard form as
\begin{align}
\bar{{\cal E}}^{\rm tot}_{A} (\hat{\rho})=&  (1-\bar{E}_{X} -\bar{E}_Y -\bar{E}_Z )\hat{\rho} \nonumber \\
&+\bar{E}_X \hat{X}_A \hat{\rho} \hat{X}_A+ \bar{E}_Y \hat{Y}_A \hat{\rho} \hat{Y}_A+
\bar{E}_Z \hat{Z}_A \hat{\rho} \hat{Z}_A,
\end{align}
the error probabilities are
\begin{align}
\bar{E}_Z=&\bar{E}_X=\frac{1}{4} -\frac{1}{4}(1-2e_{\rm m})^{2(n+1)} (1-2\bar{e}_X)^{2n}, \label{eq:EZ} \\
\bar{E}_Y=& \frac{1}{4} + \frac{1}{4} (1-2 e_{\rm m})^{2(n+1)} (1-2 \bar{e}_X)^{2n} \nonumber\\
&-\frac{1}{2} (1-2 e_{\rm m})^{2(n+1)} (1-2 \bar{e}_X)^{n}  (1-2 \bar{e}_Z)^{(2m-2)n} , \label{eq:EY} 
\end{align}
where we replace the error probability $e_{\rm d}$ of the depolarizing channel ${\cal E}$ with $e_{\rm m}$ by using Eq.~(\ref{eq:em}).
Then, note that  the average fidelity $F$ is described by
\begin{equation}
\bar{F}=1-\bar{E}_{X} -\bar{E}_Y -\bar{E}_Z.\label{eq:F} 
\end{equation}

Let us consider the scaling of the average errors $\bar{E}_{Z}$, $\bar{E}_X$, and $\bar{E}_Y$.
Suppose that, thanks to the robustness of the loss-tolerant $Z$-basis or $X$-basis measurement against depolarization (as seen in numerical examples in Sec.~\ref{se:num-rud}), $\bar{e}_Z$ and $\bar{e}_{X}$ are so small that $(1-2 \bar{e}_Z  )^{(2m-2)n}\simeq(1-2 \bar{e}_X)^{2n} \simeq1 $ holds.
In addition, we can also assume that the fidelity of the entanglement generation based on the transmission of single photons between {\it adjacent} nodes is so high as to satisfy $(1- 4 e_{\rm d}/3 )^{2(n+1)}=(1- 2 e_{\rm m} )^{2(n+1)} \simeq 1- 4 (n+1) e_{\rm m} $,
which is valid in the case where the transmission distance of photons is not long.
Then, the average errors are 
\begin{align}
\bar{E}_{Z}=\bar{E}_X\simeq \bar{E}_Y & \simeq \frac{1}{4}-\frac{1}{4} (1-2e_{\rm m})^{2(n+1)} \nonumber\\ & \simeq (n+1) e_{\rm m} = (n+1) \frac{2e_{\rm d} }{3}.
\end{align}
Since $n+1 =L/L_0$, this approximation shows that the effect of the depolarization error in the transmission from a source node to a receive node increases the error probabilities of the final entangled pair only polynomially 
with the total distance $L$.
Combined with the fact that the depolarization error in the transmission channel generally increases exponentially \cite{B98} with the channel length, 
this result implies that, thanks to the robustness of the loss-tolerant measurement for observable $\hat{Z}$ or $\hat{X}$ and the faithful transmission of a single photon over a short distance, 
our repeater protocol could suppress the effect of the channel error exponentially.

\subsection{Numerical examples}\label{se:num}
To show the polynomial scaling of our protocol explicitly, we first estimate $\bar{Q}$, $P$, $\bar{R}$,
$\bar{E}_Z$, $\bar{E}_{X}$, $\bar{E}_Y$, and $\bar{F}$ for four cases. 
Then, we provide a comparison between our protocol and the speediest protocol \cite{M12}, referencing the basic assumptions made in the updated version \cite{M12a} of Ref.~\cite{M12}.

Let us assume that photons always run in optical fibers with the transmittance $T=e^{-l/l_{\rm att}}$ for distance $l$ ($l_{\rm att}=22$ km).
In addition, we suppose that the optical fibers have small errors when they are used to connect distant repeater stations ($L_0/2$ apart), and the errors of the fiber with length $L_0/2$ can be described as an individual depolarizing channel with error probability $e_{\rm d}$.
We also assume to use single photon sources with efficiency $\eta_{\rm S}$ and photon detectors with quantum efficiency $\eta_{\rm D}$.
In the estimation of $\epsilon_0$ of Eq.~(\ref{eq:epsilon_0}), by assuming that every step in our protocol including the preparation is executed in the optical fiber, we use $P_{\tau_{\rm a} }=e^{-c \tau_{\rm a}/{l_{\rm att}}}$ with $c=2 \times 10^8$ m/s.
Suppose that the repetition rate $f$ of the slowest device among single-photon sources, photon detectors, and active-feedforward techniques is $f=100$ kHz \cite{H09}.
Then, by combining the numerical results (I) and (II) obtained in Sec.~\ref{se:num-rud} and Eqs.~(\ref{eq:tau_s}), (\ref{eq:tau_c}), (\ref{eq:P})-(\ref{eq:Q}), and (\ref{eq:EZ})-(\ref{eq:F}),
we obtain the following results under numerical calculation {\it to minimize $\bar{Q}$}:
(I) For  $L=5000$ km ($L=1000$ km), $L_0=4$ km, $e_{\rm m}=2e_{\rm d}/3=2.8 \times 10^{-5}$, $\eta_{\rm D} \eta_{\rm S}=0.95$, and $\tau_{\rm a}=150$ ns,
by choosing $m=24
$ ($m=19$), and $\{b_0,b_1,b_2\}=\{16,14,1\}$ ($\{b_0,b_1,b_2\}=\{11,11,1\}$), 
we obtain $\epsilon_0\simeq 0.20$, which presents $\bar{Q}=4.0 \times 10^7$ ($\bar{Q}=4.1 \times 10^6$),
$P=0.69$ ($P=0.58$), 
$\bar{R}=69$ kHz ($\bar{R}=58$ kHz),
$\bar{E}_{Z}=\bar{E}_X=3.5 \times 10^{-2}$ ($\bar{E}_{Z}=\bar{E}_X=8.9 \times 10^{-3}$), $\bar{E}_{Y}=3.3 \times 10^{-2}$  ($\bar{E}_{Y}=7.6 \times 10^{-3}$),
and $\bar{F}=0.90$ ($\bar{F}=0.97$).
This choice of the parameters allows us to achieve $F \ge 0.9$ as in the analysis of conventional quantum repeater schemes \cite{SSRG09}.
(II) For  $L=5000$ km ($L=1000$ km), $L_0=8$ km, $e_{\rm m}=2e_{\rm d}/3=5.6 \times 10^{-5}$, $\eta_{\rm D} \eta_{\rm S}=0.95$, and $\tau_{\rm a}=150$ ns,
by choosing $m=27
$ ($m=21$), and $\{b_0,b_1,b_2\}=\{17,28,2\}$ ($\{b_0,b_1,b_2\}=\{12,23,2\}$), 
we obtain $\epsilon_0\simeq 0.27$, which presents $\bar{Q}=7.6 \times 10^7$ ($\bar{Q}=7.3 \times 10^6$),
$P=0.65$ ($P=0.60$),
$\bar{R}=65$ kHz ($\bar{R}=60$ kHz),
$\bar{E}_{Z}=\bar{E}_X=4.0 \times 10^{-2}$ ($\bar{E}_{Z}=\bar{E}_X=1.2 \times 10^{-2}$), $\bar{E}_{Y}=3.3 \times 10^{-2}$  ($\bar{E}_{Y}=7.6 \times 10^{-3}$),
and $\bar{F}=0.89$ ($\bar{F}=0.97$).
Here we assumed that the errors of the fiber with $4$ km in the case (II) can be described as a series of two depolarizing channels for $2$-kilometer fiber in the case (I),
which is valid as long as the overall transmittance $1-\epsilon_0$ for photons is larger than the dark count probability of photon detectors \cite{BLMS00} (this is indeed the case for our repeater settings).

\subsubsection{Comparison with the speediest protocol given by Munro {\it et al.} \cite{M12}}

Let us compare our protocol with the speediest protocol given by Munro {\it et al.} \cite{M12}.
Here we refer to the data in the up-to-date version \cite{M12a} of Ref.~\cite{M12}. 

Munro {\it et al.}'s scheme uses single-photon sources and photon detectors, similarly to our proposal. Thus, in both of the protocols, suppose that the single-photon source has 97\% efficiency and the photon detector has 97\% quantum efficiency, according to \cite{M12a}.
In addition, we assume that the photon detector is the slowest device with 100~kHz repetition rate,. i.e., $f=100$~kHz. We also assume that the attenuation length of the fiber is 22~km, and 0.1\% general errors occur in the transmission of the fiber with 10 km according to \cite{M12a}. 
Although Munro {\it et al.} sometimes in their paper \cite{M12a} assume that a single photon corresponds to multiple qubits  (by using the multiple degrees of freedom such as time bin, polarization, and spatial modes), this assumption can be shared even with our protocol \cite{PM} to decrease the required photon number, and it is thus irrelevant when we make a comparison between our protocol and Munro {\it et al.}'s protocol. 
Therefore, for fairness and simplicity, here we conservatively consider that a single photon simply corresponds to a single qubit in both of the protocols. Let us consider 800 km quantum communication ($L=800$ km), according to the setting of \cite{M12a}.

\begin{description}
 \item[Performance of Munro {\it et al.}'s scheme:]

Suppose that the coupling between a single photon and a matter qubit is assumed to be 97\% \cite{M12a}. Since Table 1 of \cite{M12a} has values until $p=0.67$, we choose the distance between adjacent nodes $L_0=6.15$ km (i.e., 129 repeater nodes), because the choice actually presents $p=0.97^4\times e^{-6.15/22}=0.67$. 
We assume that the fiber with $6.15$ km includes $0.0615\%[=0.1\%\times(6.15\; {\rm km}/10 \;{\rm km})]$. Then, the Table 1 under one qubit/photon shows that the number of the matter quantum memories at each node is $13 \times 1500=19500$. Since quantum information of a single matter quantum memory is exchanged with a single photon, the number of single photons prepared at each node is also $19500$. Thus, the total number of the consumed photons for each trial is $19500\times (129+1)=2.5\times 10^6$ and the total number of matter quantum memories in repeater nodes is  $19500\times 129=2.5\times 10^6$. Since the protocol almost deterministically generates the entanglement pair with approximate fidelity $92\%[=100\%-(129+1)\times 0.0615\%]$ \cite{PM} for each trial, the rate to produce the entangled pair is about 100 kHz according to the repetition rate of the single-photon detectors.
 \item[Performance of our scheme:]
Suppose that the distance between adjacent source repeater nodes is similarly $L_0=6.15$ km (i.e., 129 source repeater nodes). This choice implies the use of fibers with (6.15/2) km, which have $0.0308\%[=0.1\%\times(6.15\;{\rm km}/2)/10\;{\rm km})]$ errors (i.e., $e_{\rm d}=0.0308$). 
By choosing $m=20$ and $\{b_0,b_1,b_2\}=\{10,20,2\}$, 
we obtain $\epsilon_0\simeq 0.25$, which presents $\bar{Q}=5.3 \times 10^6$,
$P=0.60$,
$\bar{R}=60$ kHz,
$\bar{E}_{Z}=\bar{E}_X=4.1\%$, $\bar{E}_{Y}=2.9\%$,
and $\bar{F}=89\%$,
where the number of photons prepared at each node for each trial is 24440. 
Thus, the average total number of the photons consumed until producing the entangled pair with 89\% fidelity is $5.3\times 10^6$, and the rate to produce the entangled pair is 60 kHz.
\end{description}

According to this comparison, 
our protocol is comparable with the speediest protocol of Munro {\it et al.} in the entanglement generation rate, 
although Munro {\it et al.}'s protocol uses not only single photons but also demanding matter quantum memories and both of their required numbers are in the same order of the consumed photons in our protocol.


\begin{thebibliography}{99}
\bibitem{BB84}
C.~H.~Bennett, and G.~Brassard, in {\it Proceeding of the IEEE
International Conference on Computers, Systems, and
Signal Processing, Bangalore, India} 175-179 (IEEE, New York, 1984).
\bibitem{E91}
A.~K.~Ekert, Phys. Rev. Lett.~{\bf 67}, 661 (1991).
\bibitem{B93}
 C.~H.~Bennett {\it et al.},  
Phys.~Rev.~Lett.~{\bf 70}, 1895 (1993).
\bibitem{K08}
H.~J.~Kimble, Nature {\bf 453}, 1023 (2008).
\bibitem{SSRG09}
N.~Sangouard, C.~Simon, N.~de Riedmatten, and N.~Gisin, 
Rev. Mod. Phys. {\bf 83}, 33 (2011).
\bibitem{B98}
H.~J.~Briegel, W.~D\"{u}r, J.~I.~Cirac, and P.~Zoller, 
Phys. Rev. Lett.~{\bf 81}, 5932 (1998).
\bibitem{WZ82}
W.~K.~Wootters and W.~H.~Zurek, 
Nature~{\bf 299}, 802 (1982).
\bibitem{DLCZ}
L.-M.~Duan, M.~D.~Lukin, J.~I.~Cirac, and P.~Zoller,
Nature~{\bf 414}, 413 (2001).
\bibitem{C06}
L.~Childress, J.~M.~Taylor, A.~S.~S\o rensen, and M.~D.~Lukin, 
Phys. Rev. Lett.~{\bf 96}, 070504 (2006).
\bibitem{S09}
N.~Sangouard,  R.~Dubessy, and C.~Simon, 
Phys. Rev. A {\bf 79}, 042340 (2009).
\bibitem{ATKI10}
K.~Azuma, H.~Takeda, M.~Koashi, and N.~Imoto, 
Phys. Rev. A~{\bf 85}, 062309 (2012).
\bibitem{ZDB12}
M.~Zwerger, W.~D\"{u}r, and H.~J.~Briegel,  
Phys. Rev. A {\bf 85}, 062326 (2012).
\bibitem{L12}
Y.~Li, S.~D.~Barrett, T.~M.~Stace, and S.~C.~Benjamin, 
New J. Phys.~{\bf 15}, 023012 (2013).
\bibitem{M10}
W.~J.~Munro, K.~A.~Harrison, A.~M.~Stephens, S.~J.~Devitt, and K.~Nemoto, 
Nature Photon.~{\bf 4}, 792 (2010).
\bibitem{J09}
L.~Jiang {\it et al.},
Phys. Rev. A {\bf 79}, 032325 (2009).
\bibitem{Z93}
M.~\.{Z}ukowski, A.~Zeilinger, M.~A.~Horne, and A.~K.~Ekert, 
Phys. Rev. Lett.~{\bf 71}, 4287 (1993).
\bibitem{W02}
E.~Waks, A.~Zeevi, and Y.~Yamamoto,
Phys. Rev. A {\bf 65}, 052310 (2002).
\bibitem{R09}
M.~Razavi, M. Piani, and N.~L\"utkenhaus,
Phys. Rev. A {\bf 80}, 032301 (2009).
\bibitem{KLM01}
E.~Knill, R. ~Laflamme, and G.~J.~Milburn, 
Nature {\bf 409}, 46 (2001).
\bibitem{D00}
D.~P.~DiVincenzo,   
quant-ph/0002077.
\bibitem{G12}
A.~Grudka {\em et al.},
arXiv:1202.1016.
\bibitem{M12}
W.~J.~Munro, A.~M.~Stephens, S.~J.~Devitt, K.~A.~Harrison, and K.~Nemoto, 
Nature Photon.~{\bf 6}, 777 (2012).
\bibitem{L10}
T.~D.~Ladd {\it et al.},
Nature {\bf 464}, 45 (2010).
\bibitem{MK13}
C.~Monroe and J.~Kim, Science {\bf 339},1164 (2013).
\bibitem{S10}
C.~Simon {\em et al.},
Eur. Phys. J. D {\bf 58}, 1 (2010).
\bibitem{P07}
R.~Prevedel {\em et al.},
Nature {\bf 445}, 65 (2007).
\bibitem{VBR06}
M.~Varnava, D.~E.~Browne, and T.~Rudolph,
Phys.~Rev.~Lett.~{\bf 97}, 120501 (2006).
\bibitem{L09}
N.~H.~Lindner and T.~Rudolph,
Phys.~Rev.~Lett.~{\bf 103}, 113602 (2009).
\bibitem{T05}
S.~Tanzilli {\it et al.},
Nature {\bf 437}, 116 (2005).
\bibitem{I11}
R.~Ikuta {\it et al.},
Nature Commun. {\bf 2}, 537 (2011).
\bibitem{GC99}
D.~Gottesman and I.~L.~Chuang, 
Nature {\bf 402}, 390 (1999).
\bibitem{RB01}
R.~Raussendorf and H.~J.~Briegel,
Phys. Rev. Lett.~{\bf 86}, 5188 (2000).
\bibitem{LCQ12}
H.-K.~Lo, M.~Curty, and B.~Qi,
Phys. Rev. Lett. {\bf 108}, 130503 (2012).
\bibitem{W94}
H.~Weinfurter, 
Europhys. Lett. {\bf 25}, 559 (1994).
\bibitem{V03}
L.~Vaidman, 
Phys. Rev. Lett. {\bf 90}, 010402 (2003).
\bibitem{LL11}
H.-K.~Lau and H.-K.~Lo, 
Phys. Rev. A {\bf 83}, 012322 (2011).
\bibitem{B11}
H.~Buhrman {\it et al}., 
{\it Advances in Cryptology -- CRYPTO 2011} {\bf  6841}, 429 (Springer Berlin Heidelberg, 2011). 
\bibitem{B13}
M.~A.~Broome {\it et al.},
Science {\bf 339}, 779 (2013).
\bibitem{S13}
J.~B.~Spring {\it et al.},
Science {\bf 339}, 798 (2013).
\bibitem{VBR08}
M.~Varnava, D.~E.~Browne, and T.~Rudolph,
Phys.~Rev.~Lett.~{\bf 100}, 060502 (2008).
\bibitem{VBR07}
M.~Varnava, D.~E.~Browne, and T.~Rudolph, 
New~J.~Phys. {\bf 9}, 203 (2007).
\bibitem{GK}
D.~Gottesman, 
{\it In Group22: 
Proceedings of the XXII International Colloquium on Group Theoretical Methods in Physics}, 
32-43 (Cambridge, MA, International Press, 1999).
\bibitem{HDERVB}
M.~Hein, {\em et al.},
quant-ph/0602096.
\bibitem{CCWD06}
Q.~Chen, J.~Cheng, K.-L.~Wang, and J.~Du, 
Phys.~Rev.~A~{\bf 73}, 012303 (2006).
\bibitem{FT10}
K.~Fujii and Y.~Tokunaga,
Phys.~Rev.~Lett. {\bf 105}, 250503 (2010).
\bibitem{KWD03}
P.~Kok, C.~P.~Williams, and J.~P.~Dowling, 
Phys. Rev. A {\bf 68}, 022301 (2003).
\bibitem{L06}
P.~van Loock {\it et al.},
Phys. Rev. Lett.~{\bf 96}, 240501 (2006).
\bibitem{La06}
T.~D.~Ladd {\it et al.},
New J. Phys.~{\bf 8}, 184 (2006).
\bibitem{S07}
C.~Simon~{\it et al.},
Phys. Rev. Lett. {\bf  98}, 190503 (2007).
\bibitem{L08}
P.~van Loock, N.~L\"utkenhaus, W. J.~Munro, and K.~Nemoto,
Phys. Rev. A {\bf 78}, 062319 (2008).
\bibitem{M08}
W.~J.~Munro, R.~Van Meter, S.~G.~R.~Louis, and K.~Nemoto,
Phys. Rev. Lett. {\bf 101}, 040502 (2008).
\bibitem{A09}
K.~Azuma, {\it et al.},
Phys. Rev. A {\bf 80}, 060303(R) (2009).
\bibitem{A10}
K.~Azuma, N.~Sota, M.~Koashi, and N.~Imoto, 
Phys. Rev. A {\bf 81}, 022325 (2010).
\bibitem{C10}
S.~R.~Clark, A.~J.~Connor, D.~Jaksch, and S.~Popescu, 
New J. Phys. {\bf 12}, 083034 (2010).
\bibitem{KMS11}
A.~Kent, W.~J.~Munro, and T.~P.~Spiller,
Phys. Rev. A {\bf 84}, 012326 (2011).
\bibitem{H09}
R.~H.~Hadfield, 
Nature Photon. {\bf 3}, 696 (2009).
\bibitem{BLMS00}
G.~Brassard, N.~L\"utkenhaus, T.~Mor, and B.~C.~Sanders, 
Phys. Rev. Lett. {\bf 85}, 1330 (2000).
\bibitem{M12a}
W.~J.~Munro, A.~M.~Stephens, S.~J.~Devitt, K.~A.~Harrison, and K.~Nemoto, 
arXiv:1306.4137.
\bibitem{PM}
Private communication with W.~J.~Munro.
\end{thebibliography}
\end{document}